\documentclass[pra,aps,amsmath,amssymb,superscriptaddress,twocolumn, longbibliography]{revtex4-1}
\usepackage{graphicx}
\usepackage{color,outlines}
\usepackage[english]{babel}
\usepackage{amsthm}
\usepackage[sc,osf]{mathpazo}
\usepackage{hyperref}
\usepackage{physics}
\usepackage{outlines}
\usepackage{tabularx}
\usepackage{mathtools}
\usepackage{lipsum}

\definecolor{green2}{RGB}{0,100,0}

\usepackage{tikz, pgfplots}
\usetikzlibrary{positioning}

\begin{document}
	
	\title{Evaluating quantum circuits in the reservoir computing paradigm}
	
	\author{Gaurav Rudra Malik}
	\email{gauravrudramalik.rs.phy22@itbhu.ac.in}
	\affiliation{Department of Physics, Indian Institute of Technology (Banaras Hindu University), Varanasi, India~221005} 
	
	\author{Amit Kumar Jaiswal}
	\email{amit.chr@iitbhu.ac.in}
	\affiliation{Jay Chaudhry Software Innovation Centre, Department of Computer Science and Engineering, Indian Institute of Technology (Banaras Hindu University), Varanasi, India~221005} 
	
	\author{S. Aravinda}
	\email{aravinda@iittp.ac.in}
	\affiliation{Department of Physics, Indian Institute of Technology Tirupati, Tirupati, India~517619} 
	
	\author{Sunil Kumar Mishra}
	\email{sunilkm.app@iitbhu.ac.in}
	\affiliation{Department of Physics, Indian Institute of Technology (Banaras Hindu University), Varanasi, India~221005}

	\begin{abstract}
		Reservoir computing is a framework for learning temporal correlations in time series data by leveraging the intrinsic dynamics of an underlying physical system. In quantum implementations, the reservoir is commonly realised through the ergodic dynamics of nonintegrable Hamiltonian models. In our present work, we investigate the relation between reservoir performance and underlying dynamics in an analogous setting wherein a structured brickwall quantum circuit functions as the reservoir. We focus on the manner in which the choice of constituent two qubit gates influences task performance  correlating the same with known results on the dynamical nature of the circuit in question. Our primary analysis involves the class of dual-unitary circuits, for which features governing the global dynamics are tunable and exactly solvable. We further study a class of circuits constructed out of non-random two qubit gates satisfying a specific solvability condition under the Cartan decomposition, for which the second moment spectrum can be determined analytically. For sufficiently large entangling power of gates, the said circuit converges to unitary designs faster than an equivalent brickwall arrangement of Haar random two qubit gates. In both cases considered, the dynamics are respectively characterised by analytically accessible exact measures of system level complexity. Our results show that reservoir performance is strongly influenced by the choice of local two-qubit operations, and that structured quantum circuits can function as effective reservoirs, where the peak performance can be accessed in relatively modest circuit depths when compared to conventional Hamiltonian reservoirs.	
	\end{abstract}

	\maketitle
	\section{Introduction}
	\label{Introduction}
	
	In recent decades the techniques involving deep neural networks have had a profound impact on the application of machine learning \cite{nielsenneural}. In quantum computing, combining neural networks with the prevalent circuit architectures has lead to established pathways in the area of Quantum Machine Learning, making use of techniques such as variational algorithms and parametrised circuits \cite{ML1, ML2, ML3, ML4, ML5, ML6, ML7, ML8, ML9, ML10, ML11, ML12, ML13, ML14}. An important application with neural networks is the ability to analyse and make predictions over time-series data, where existing architectures like Recurrent Neural Networks (RNN) and Long Short-Term Memory (LSTM)  have proved to be effective, although resource intensive. An alternate paradigm, termed as reservoir computing, has been explored for time-series prediction which avoids the associated computational overhead by transferring the input to a randomly initialised RNN, from which meaningful predictions can be made upon training only the final output layer. The randomly initialised RNN does not require optimising any other of the associated weights and biases. For the approach to be effective it is important that the underlying RNN has specific ergodic characteristics in order to facilitate the mapping to a higher dimensional space, while retaining the stability of operation \cite{Stepney2024, CRQC1,CRQC2,CRQC3,CRQC4,CRQC5,CRQC6}. 
	
	To replicate reservoir computing in a quantum setting, it is required to have a physical model that shares the ergodic nature of a randomly initialised RNN \cite{Fuji_harnessing_disorder}. This can be achieved primarily using the unitary evolution operator corresponding to ergodic models from many-body physics which have an inherent thermalising nature \cite{QRC2}. In this direction several important studies have been made in the present literature, highlighting the applicability of this approach \cite{QRC1, QRC3, QRC4, QRC_Exp1, QRC_Finance}. An important consideration is how time-series prediction performance depends on the degree of ergodicity within the underlying dynamics. Optimal performance is typically observed near the 'edge-of-chaos' regime, where ergodicity is balanced against dynamical stability \cite{kobayashi_edgeofchaos}. With increasing the degree of ergodicity the mapping of input data to a higher dimensional feature space is enhanced, which can improve task performance. However, stronger chaotic dynamics reduce stability, where nearby input states may evolve into increasingly separated representations. Recent works have also shown the experimental realisation of quantum reservoirs \cite{QRC_Exp1}, with the observed performance improving on benchmark models \cite{QRC_Finance}. In all above examples, the time series prediction task is formulated as a supervised machine learning task, where the model is trained on a series of input and output pairs, while being evaluated on a separate test dataset. Prediction accuracy is quantified by comparing the model output with the corresponding target values over the entire test dataset.
	
	In our present work, we study task performance in quantum reservoir computing, using the class of structured quantum circuits as the underlying reservoir. The central objective of our present study is to determine how the choice of constituent gates in a brickwall circuit influences task performance, given that it is already known to significantly modify the overall dynamical behaviour of the circuit \cite{Operator_Spreading_nahum2, Piroli_Bertini, Operator_Spreading_Claeys, Circuit_quasiparticles, Sourav_MIPT, more_global_randomness}. The reservoir in consideration, i.e. the aforementioned brickwall circuit, is known to have features associated with thermalising dynamics \cite{Fisher_RQC}. For example, even being composed of generic two-qubit gates the circuit has light-cone velocity equal to unity, where the influence of an initial operator spreads at the maximum rate of a single site per time step. Moreover, an absence of conservation laws and an integrable structure prevents additional dynamical constraints from appearing \cite{Chalker_minimal_models, Operator_spreading_nahum1}. This is accompanied by a rapid growth of OTOCs and a linear increase in entanglement, albeit at a sub-maximal rate \cite{Rohit_PRB}. Replacing the two-qubit unitaries forming the circuit with special structured examples like dual-unitary gates enables us to access dynamics where the said parameters have their maximal values, leading to the class of dual-unitary circuits being referred to as minimal models of maximal many-body chaos \cite{Claeys_Dual, Dual_Unitary1, Dual_Unitary2, Bertini2020OperatorEntanglementI, Bertini2020OperatorEntanglementII, Dual_Unitary_Review}. 
	
	Another desirable feature associated with the class of dual-unitary circuits is the emergence of an ergodicity-hierarchy characterised by the entangling power $e_P(U)$ of the involved quantum gates. Increasing $e_P(U)$ leads to the build-up of ergodic behaviour and separates mixing dynamics from trivial examples such as the evolution generated by a sequence of SWAP gates, due to their non-entangling nature irrespective of being dual-unitary \cite{Aravinda_Sir_PRR}. In our work we specifically make use of this tunable nature in evaluating the reservoir for task performance, over synthetically generated time series data. The overlap between ergodic dynamics and reservoir computing is further established by diagnostics used to measure information spread being reliable indicators for task performance \cite{Wisniacki_Krylov_first, Cindrak_Krylov_Expressivity, Cindrak_Krylov_Observability}. 
	
	Our results convey that brickwall quantum circuits maybe used effectively for reservoir computing applications, wherein their structural nature leads to a reduced circuit depth for attaining the maximal spread of input data. Against a bonafide minimal example of a brickwall circuit involving two-qubit Haar random gates there is a reduction in computational overhead which follows from the requirement of single qubit Haar random local operators for the case of structured circuits like dual-unitary models\cite{Efficiency_RQC, design1, design2, Haferkamp_design, frame_potential2, design_otoc}.
	
	The manuscript is organised as follows: We begin with describing the steps involved in using circuit reservoirs in Section \ref{Brickwall_Reservoirs}, followed by a description of dual-unitary circuits and associated mixing dynamics in Section \ref{Duals_section}. We then describe our prediction tasks based on synthetic datasets introduced in Section \ref{datasets}, using which reservoir performance is benchmarked in Section \ref{Benchmarking}. We also discuss the role of multiplexing in Section \ref{Multiplexing}. The class of solvable circuits in context of reservoirs is described in Section \ref{Special}. We summarise our results and conclude with our work in Section \ref{Summary}.
	
	\section{Reservoir Computing with Brickwall Quantum Circuits}
	\label{Brickwall_Reservoirs}
	\begin{figure*}
		\centering
		\usetikzlibrary {decorations.pathmorphing, decorations.pathreplacing, decorations.shapes, arrows.meta}
		\resizebox{\textwidth}{!}{%
			\begin{tikzpicture}
				\foreach \i in {0,...,5}
				{
					\draw[thick] (7, 2*\i) -- (73, 2*\i);
				}
				
				\draw[thick, dashed] (8.9,-2) -- (8.9,12);
				\draw[thick, dashed] (65.9,-2) -- (65.9,12);
				\draw[thick, dashed] (9.1,-2) -- (9.1,12);
				\draw[thick, dashed] (66.1,-2) -- (66.1,12);
				
				\draw [thick, fill=white, dashed] (68.5, -1.0) rectangle (72.5, 11.5);
				\draw [thick, fill=white, dashed] (11.5, -1.0) rectangle (15.5, 11.5);
				
				\draw [thick,fill=white,rounded corners=2.2pt] (69, 9.0) rectangle (72,11);
				\node [font = \Huge] at  (70.5,10) {Inj.};
				\node [font = \Huge] at  (70.5,12) {Input: $s_{k+1}$};
				
				\draw [thick,fill=white,rounded corners=2.2pt] (69, -0.5) rectangle (72,8.5);
				\node [font = \Huge] at  (70.5,4.5) {$\tr_1(\rho)$};
				
				\draw [thick,fill=white,rounded corners=2.2pt] (12, -0.5) rectangle (15,8.5);
				\node [font = \Huge] at  (13.5,4.5) {$\tr_1(\rho)$};
				\draw [thick,fill=white,rounded corners=2.2pt] (12, 9.0) rectangle (15,11);
				\node [font = \Huge] at  (13.5,10) {Inj.};
				\node [font = \Huge] at  (13.5,12) {Input: $s_k$};
				
				\draw [thick,fill=white,rounded corners=2.2pt] (19, -1.0) rectangle (23,11);
				\node [font = \Huge] at  (21,4.5) {$U_{\text{res}}$};
				\draw[-{Latex[length=9mm]}] (26,10) -- (26,-2);
				\node[font=\Huge] at (26.65,-5.5) {$\begin{pmatrix} z_1^1 \\ z_2^1 \\ \vdots \\ z_5^1 \\ z_6^1 \end{pmatrix}_{n \times 1}$};
				
				\foreach \i in {0,...,5}
				{
					\draw[thick, fill=black] (26, 2*\i) circle (0.25);
				}
				\draw [thick,fill=white,rounded corners=2.2pt] (29, -1.0) rectangle (33,11);
				\node [font = \Huge] at  (31,4.5) {$U_{\text{res}}$};
				\draw[-{Latex[length=9mm]}] (36,10) -- (36,-2);
				\node[font=\Huge] at (36.65,-5.5) {$\begin{pmatrix} z_1^1 \\ z_2^1 \\ \vdots \\ z_5^2 \\ z_6^2 \end{pmatrix}_{2n \times 1}$};
				
				\foreach \i in {0,...,5}
				{
					\draw[thick, fill=black] (36, 2*\i) circle (0.25);
				}
				\draw [thick,fill=white,rounded corners=2.2pt] (39, -1.0) rectangle (43,11);
				
				\node [font = \Huge] at  (41,4.5) {$U_{\text{res}}$};
				\draw[-{Latex[length=9mm]}] (46,10) -- (46,-2);
				\node[font=\Huge] at (46.65,-5.5) {$\begin{pmatrix} z_1^1 \\ z_2^1 \\ \vdots \\ z_5^3 \\ z_6^3 \end{pmatrix}_{3n \times 1}$};
				\foreach \i in {0,...,5}
				{
					\draw[thick, fill=black] (46, 2*\i) circle (0.25);
				}
				
				\draw [white,fill=white] (48, -1.0) rectangle (54,11);
				\node [font = \Huge, align = center] at  (51,4.5) {Repeating for \\ V steps};
				
				\draw [thick,fill=white,rounded corners=2.2pt] (56, -1.0) rectangle (60,11);
				\node [font = \Huge] at  (58,4.5) {$U_{\text{res}}$};
				\draw[-{Latex[length=9mm]}] (63,10) -- (63,-2);
				\node[font=\Huge] at (63.65,-5.5) {$\begin{pmatrix} z_1^1 \\ z_2^1 \\ \vdots \\ z_5^v \\ z_6^v \end{pmatrix}_{nv \times 1}$};
				\foreach \i in {0,...,5}
				{	
					\draw[thick, fill=black] (63, 2*\i) circle (0.25);
				}
				
			\end{tikzpicture}
		}
		\caption{A diagrammatic description of the quantum reservoir computing approach, implemented using a $6$ qubit quantum circuit. The time series input $\{ s_k \}_{k=1}^{L}$ is sequentially injected into the state of the system, as shown for the case $s_k$. The state therefore contains information of the given time-series, and is subsequently acted upon by the reservoir unitary a total of $V$ times, followed by projective measurements. The data extracted is collected, leading to a $NV \times 1$ vector for each data-point. The process is repeated for successive data-points leading to the final $Z$ matrix having dimensions $NV \times L$.}
		\label{Figure_Reservoir_Schematics}
	\end{figure*}
	The central object that we study in this work is a quantum circuit constructed by staggered layers of two qubit gates arranged in a brickwall pattern that functions as a quantum reservoir. We shall begin with outlining the general method associated with quantum reservoir computing that is used for time-series prediction as outlined in Ref. \cite{Fuji_harnessing_disorder}, and its adaptation for the circuit case. The method used is simple from an implementation standpoint, although at present there exist several improvements over the original protocol, specifically introducing features such as feedback \cite{Kobayashi_feedback} and measurement back-action \cite{Dissipation_resource, backaction_mujal, Eisert_QRC}. Since our work specifically considers the role of a reservoir and its associated ergodic nature towards task performance, it is better to study a simpler implementation without additional features known to enhance prediction accuracy.
	
	For our present implementation, we generate a sequence of synthetic data using established methods present in computer science literature \cite{Benchmarks}. The sequential time-series data is given as $\{s_k,y_k\}_{k = 1}^{L}$, where $s_k$ and $y_k$ refer to the input and output values related by an unknown non-linear function, with $L$ being the length of the total segment. The objective of reservoir implementation is to learn features of the unknown function and subsequently map an input to its most probable output while minimising the errors over an entire dataset. The input sequence \(\{s_k\}\), is normalised between $[0,1)$ and is injected at discrete times into the first qubit of the quantum circuit. For this, at each step $k$, we initialise a pure state given by:
	\begin{equation}
		\ket{\psi_{s_k}}
		=
		\sqrt{1-s_k}\,\ket{0}
		+
		\sqrt{s_k}\,\ket{1},
	\end{equation}
	so that the corresponding single-qubit density matrix is
	\begin{equation}
		\rho_{s_k}=\ket{\psi_{s_k}}\bra{\psi_{s_k}}.
	\end{equation}
	The full reservoir state is then updated through the completely positive trace-preserving map:
	\begin{equation}
		\rho \;\longrightarrow\; \rho_{s_k}\otimes \mathrm{Tr}_{1}[\rho],
	\end{equation}
	where \(\mathrm{Tr}_1\) denotes the partial trace over the first qubit.
	
	This state is then time evolved by the unitary reservoir operator $U_{res}$, which in our case is defined by the single time-step of the brickwall circuit. The details of $U_{res}$ follow the present discussion. Between the injection of any two data points $s_k$ and $s_{k+1}$, the reservoir state $\rho$ is time evolved a total of $V$ times, a process which is referred to as \emph{multiplexing}. After each application of the reservoir unitary $U_{res}$, we evaluate the expectation values for the $N$ single qubit operators $Z_i$ for each of the $i$ qubits ($i = 0,1,...,N$). This leads to the values $z_i^j$, following the $j$th application of the reservoir unitary on the $i$th qubit. Therefore, we have:

	\begin{equation}
		z_i^j
		=
		\mathrm{Tr}\!\left[\frac{(I+Z_i)}{2} U_{res}^j \rho {U^j}^\dagger_{res} \right]
		=
		\frac{\langle Z_i^j\rangle +1}{2}.
	\end{equation}
	where, $\langle Z_i^j\rangle = \Tr(Z_i U_{res}^j \rho {U^j}^\dagger_{res})$. Each of these extracted values for $z_i^j$ and stacked upon each other to create a column vector of dimension $NV \times 1$, before the exact same process is repeated for the next input state $s_{k+1}$. The outline of this process is shown in Fig. \ref{Figure_Reservoir_Schematics}.

	At this point, the numerical trace evaluation used to obtain \(z_i^j\) should be distinguished from the corresponding measurement procedure in a physical quantum reservoir implementation. The latter requires a series of projective measurements over an ensemble of identically prepared circuits, where each member is discarded once measured. The ensemble averaging then gives us the required expectation values. Thus, by simply taking the expectation values via by tracing an operator with the density matrix of the state implicitly ignores the effect of measurement backaction that leads to the collapse of the original wavefunction. This is also an assumption taken in Ref. \cite{Fuji_harnessing_disorder}, and remains valid in a numerical realisation of the reservoir computing setup. This also clarifies that following measurement, we are still working with the state evolved from the successive actions of the reservoir unitary $U_{res}$, instead of some generic eigenstate of the measured operator.
		
		The notion of multiplexing is important as it generates a large collection of data points to train the output of the model, for each input $s_k$. The process we describe is analogous to time multiplexing carried out in Ref. \cite{Fuji_harnessing_disorder} for Hamiltonian evolution, where, instead of sampling once , the interval \(\tau\) is divided into \(V\) equal subintervals and with the signal being recorded at each intermediate time step given by:
		\begin{equation}
			t = k\tau + v\frac{\tau}{V},
			\qquad v=1,\dots,V.
		\end{equation}
		At each subinterval, the measured signal from qubit \(i\), for the operator $Z_i$ is given as:
		\begin{equation}
			x_i\!\left(k\tau+v\frac{\tau}{V}\right)
			=
			\mathrm{Tr}\!\left[\frac{I+Z_i}{2}\,\rho\!\left(k\tau+v\frac{\tau}{V}\right)\right] = x_{iv}.
		\end{equation}
		Here, we have introduced the term $x_{iv}$ as a shorthand notation where index  \(i=1,\dots,N\) labels the qubit and index \(v\) references the subinterval. These intermediate samples $x_{iv}$ are defined as virtual nodes. Therefore, although the physical system has only \(N\) measured qubits, the multiplexed readout yields \(NV\) computational nodes per input step. These time multiplexed signals therefore generate a larger set of features for the classical postprocessing, which is described later in this section.
	
	In this way, via multiplexing, one input injection followed by one interval of quantum evolution produces an entire block of \(NV\) measured features for the linear readout layer leading to additional effective reservoir nodes. This is especially viable in a quantum setup, as the physically measured observables are only the \(N\) single qubit \(Z_i\) values, while the full reservoir state lives in an exponentially large operator space. By sampling the transient dynamics at \(V\) intermediate times, the protocol extracts a richer set of features from the same underlying quantum evolution without increasing the number of physical qubits. We may therefore emphasize that the total number of computational nodes entering the learning stage is \(NV\), instead of \(N\), for each input point $s_k$. Thus, the output $z_i^j$ extracted within the approach of a circuit reservoir corresponds to the output $x_{iv}$ described for the Hamiltonian reservoir. Note that for the latter, both $\tau$ and $V$ are operational parameters, which tend to affect the performance for a given time series. In the case of a circuit reservoir, the time step is defined by the geometric construction of the brickwall circuit to be unity and cannot be altered. Therefore, multiplexing involves repeated application of the reservoir unitary followed by evaluating the expectation values, instead of the same at intermediate times as in the case of Hamiltonian evolution.
	
	The entire process when repeated for all of the input parameters $s_k$, yields $L$ vectors of dimension $NV \time 1$. All of these are collected together to form the $Z$ matrix of order $NV \times L$, using which we can find the optimal weight parameters to make the time series predictions leading to a set of predicted outputs $\{\tilde{y}_k \}$ against the provided inputs $\{s_k\}$. For this we use appropriate regression methods and it constitutes the classical postprocessing step. For this purpose, the total dataset of length $L$ is divided into a training and an evaluation part. Specifically, \(80\%\) of the total $L$ points $\{ s_k, y_k \}$ are used for training ($L_t$) while the remaining \(20\%\) are kept for the purpose of evaluation ($L_e$). The $Z$ matrix is also correspondingly split, leading us with the following matrices:
	\begin{equation}
		Z_{\mathrm{train}} = Z_{NV \times L_{t}},
		\qquad
		y_{\mathrm{train}} = \{y_k\}_{k = 1}^{L_{t}},
	\end{equation}
	\begin{equation}
		Z_{\mathrm{eval}} = Z_{NV \times L_{e}},
		\qquad
		y_{\mathrm{eval}} = \{y_k\}_{k = L_{t}}^{L}.
	\end{equation}
	The linear readout weights $w$ are obtained only from the training data by solving the least squares problem
	\begin{equation}
		Z_{\mathrm{train}}^{T} w \approx y_{\mathrm{train}}.
	\end{equation}
	For inverting the above equation, it is most straightforward to evaluate the Moore-Penrose pseudoinverse and is commonly performed step in regression problems. For this, we calculate:
	\begin{equation}
		Z_{\mathrm{train}}^{+} = \mathrm{pinv}(Z_{\mathrm{train}}^T),
	\end{equation}
	and compute the optimal linear weights as
	\begin{equation}
		w = Z_{\mathrm{train}}^{+} y_{\mathrm{train}}.
	\end{equation}
	This gives the minimum-norm least squares solution for the linear regression problem. Finally, the trained weights are applied to the held out evaluation data to obtain the predictions,
	\begin{equation}
		y_{\mathrm{pred}} = Z_{\mathrm{eval}}^T w.
	\end{equation}
	Thus, the procedure consists of organising the reservoir signals into a regression matrix, splitting the data into training and testing segments, followed by fitting the linear readout through the pseudoinverse on the training set alone. A similar method based on weak measurement tomography also exists which considers the extracted data to characterise quantum dynamics \cite{State_reconstruction_krylov}. Finally the predictive performance is assessed on the unseen evaluation data. As the measure of task performance, we consider the metric of mean squared error (MSE), defined as:
	\begin{equation}
		MSE = \frac{1}{L_e} \sum_{i = 1}^{L_e} (y_{\text{pred}}^i - y_{\text{eval}}^{i})^2
	\end{equation}
	As the predictive performance leads to convergence such that the MSE is of the order $\sim 10^{-5}$, the quantity that is reported in our work is that of $log_{10}(MSE)$, for which a lower value indicates better task performance. 
	
	\section{Dual Unitarity, Mixing Dynamics and Krylov Space Analysis}
	\label{Duals_section}
	
	In this work, we are primarily interested in studying the effectiveness of dual unitary quantum circuits as reservoirs. In order to satisfy the dual unitary condition, a two qubit gate remains as a unitary matrix following the realignment operation. For a brickwall circuit constructed out of dual unitary gates, the correlation function takes on a solvable form and wherein we can investigate the role between ergodicity and task performance (see Appendix \ref{Mixing_Rate} and Refs. \cite{Dual_Unitary1, Aravinda_Sir_PRR, Dual_Unitary_Review}).
	
	For analysing brickwall circuits, it is convenient to introduce the graphical notation used to describe quantum circuits having the said structure. Here, the local 2 qubit gate is represented as a block, along with its hermitian conjugate in a different colour:
	\begin{equation}
		\begin{tikzpicture}[baseline=(current  bounding  box.center), scale=0.8]
			\draw[thick] (-4.25,0.5) -- (-3.25,-0.5);
			\draw[thick] (-4.25,-0.5) -- (-3.25,0.5);
			\draw[thick, fill=purple, rounded corners=2pt] (-4,0.25) rectangle (-3.5,-0.25);
			\node at (-4.75,0.05){$\hat{U}=$};
			\node at (-3.,-0.075){,};
		\end{tikzpicture}
		\qquad 
		\qquad
		\begin{tikzpicture}[baseline=(current  bounding  box.center), scale=0.8]
			\draw[thick] (-1.25,0.5) -- (-.25,-0.5);
			\draw[thick] (-1.25,-0.5) -- (-.25,0.5);
			\draw[thick, fill=teal, rounded corners=2pt] (-1,0.25) rectangle (-0.5,-0.25);
			\node at (-1.75,0.075){$\hat{U}^\dag=$};
			\node at (0,-0.075){.};
		\end{tikzpicture}
		\label{U_tensor}
	\end{equation}
	The graphical notation is frequently used for generic brickwall circuits as well, not necessarily satisfying the conditions of dual unitarity \cite{Fisher_RQC}.
	
	Similar notation was used in our previous analysis, relating entanglement growth and mixing in dual unitary circuits \cite{literature1}. Although the following framework is not a new result in the present work, the mixing rate remains relevant, as it provides the dynamical basis for characterising the circuit reservoir. For a brickwall arrangement, the nature of single block $U$ that constructs the circuit also defines several associated properties of the global dynamics \cite{Operator_Spreading_nahum2, Piroli_Bertini, Operator_Spreading_Claeys, Circuit_quasiparticles, Sourav_MIPT, more_global_randomness}, irrespective of dual unitarity. An important example is that of mixing rate which is determined by the entangling power $e_P(U)$ \cite{Aravinda_Sir_PRR} of individual gates constructing the brickwall circuit when the dual unitary condition is satisfied. The metric $e_P(U)$ quantifies average entanglement generated for an ensemble of product states \cite{Zanardi_entangling_power} and remains invariant under the action of local, single qubit unitaries.
	
	The unitary evolution via the reservoir $U_{res}$ is taken as a single layer of the brickwall circuit, which has the following form, for an even number of qubits,
	\begin{equation}
		U_{res} =
		\Bigg[ \bigotimes\limits_{i \in \mathcal{Z}_{\text{even}}} \hat{U}^{i,i+1} \Bigg]
		\cdot
		\Bigg[ \bigotimes\limits_{j \in \mathcal{Z}_{\text{odd}}} \hat{U}^{j,j+1} \Bigg].
	\end{equation}
	Here $\hat{U}^{pq}$ represents the operator $\hat{U}$ acting on the $p th$ and $q th$ lattice site. Using the graphical notation specified in Eq. \ref{U_tensor} we therefore have $U_{res}$ for $N = 6$ with the lattice points labelled as $\{ 0, 1, ..., 5\}$. The setup also assumes periodic boundary conditions, indicated by the contraction of rightmost leg with the index corresponding to the $0$th lattice point.
	\begin{equation}
		U_{res} =
		\begin{tikzpicture}[baseline=(current  bounding  box.center), scale=0.3]
			
			\draw[thick] (11.5,1.0) -- (9.5,3.0);
			\draw[thick] (-0.5,1.0) -- (1.5,-1.0);
			\draw[thick,dotted] (11.5,1.0) -- (-0.5,1.0);
			
			\foreach \i in {0,1}
			{
				\draw[thick] (5.5 + 4*\i, -1.0) -- (1.5 + 4*\i,3.0);
			}	
			
			\foreach \i in {0,...,2}
			{
				\draw[thick] (-0.5 + 4*\i, -1.0) -- (3.5 + 4*\i,3.0);	
				
				\draw [thick,fill=purple,rounded corners=2.2pt] (4*\i,-0.5) rectangle (1 + 4*\i,0.5);
				
				\draw [thick,fill=purple,rounded corners=2.2pt] (2 + 4*\i,1.5) rectangle (3 + 4*\i,2.5);

			}
			
			\node at (-0.5 ,-2.0) {0};
			\node at ( 3.5 ,-2.0) {2};
			\node at ( 7.5 ,-2.0) {4};
			
			\node at (1.5 ,-2.0) {1};
			\node at (5.5 ,-2.0) {3};
			\node at (9.5 ,-2.0) {5};
			
			
		\end{tikzpicture}
		\label{operator_tensor}
	\end{equation}
	For a total of $t$ time steps, the total time evolution operator is thereby given as: $\mathbb{U}(t) = U_{res}^t$, although in practice, we only use $U_{res}$ at each step, as unitary evolution is followed by taking expectation values in the reservoir computing process.  
	
	The local single qubit unitaries play an important role in determining the dynamical properties of the system. While entangling power is local unitary invariant, taking an ensemble of operators $U' = (u_1 \otimes u_2) U (v_1 \otimes v_2)$ leads to a range of dynamical behaviours, quantified by the varying mixing rate of the circuit \cite{Aravinda_Sir_PRR}. Here, it can be observed that $U'$ remains dual unitary provided $U$ belongs to the same class. Note that considering $U'$ as a building block for $U_{res}$ implies that each brick within the circuit in Eq. \ref{operator_tensor} is the same, and is given by $U'$ for an initially sampled set of $\{u_i,v_i\}$ from single qubit Haar random unitaries. Thus, the reservoir operation $U_{res}$ has a Floquet structure instead of a random quantum circuit setup where the local single qubit unitaries are independently sampled at each spacetime point. The role of local operators in defining the mixing rate of the circuit is described further in Appendix \ref{Mixing_Rate}.
	
	In order to resolve the variability arising out of local unitaries, we choose those single qubit Haar operators which maximise the mixing rate of the circuit. Thus, for a given value of entangling power, the gate used is the one with the maximal mixing rate, and this particular gate is repeated throughout to construct the Floquet structure. For this purpose the local random unitaries $\{u_i,v_i\}$ sampled have the form:
	\begin{equation}
		w(\phi, \psi) = \frac{1}{\sqrt{2}}\begin{pmatrix}
			e^{i\frac{\phi}{2}} & - e^{-i\frac{\psi}{2}} \\
			e^{-i\frac{\psi}{2}} & e^{-i\frac{\phi}{2}}
		\end{pmatrix}
		\label{w_condition}
	\end{equation} 
	where the maximal mixing rate for a fixed entangling power can be analytically solved. The mixing rate ($\mu$) follows from the magnitude of the largest non-trivial eigenvalue $\lambda_1$, and is given as $\mu_1 = -\ln(\lambda_1)$. The maximal mixing rate iver all local random unitaries is related to the entangling power $e_P(U)$, and is given as \cite{Aravinda_Sir_PRR}:
	\begin{equation}
		\text{max}_{u_i,v_i} \Big( \mu_1 \Big) = - \frac{1}{3} \ln \Bigg[1 - \frac{e_P(U)}{e^{max}_P(U)} \Bigg],
		\label{max_mixing}
	\end{equation}

	The form specified in Eq. \ref{w_condition} is for an arbitrary 2 qubit gate given as $u(\theta, \phi, \psi) \in SU(2)$, with $\theta = \pi/2$. It has also been shown that maximising the mixing over the subset $w$ of all possible gates, gives the overall maximal mixing rate, following Eq. \ref{max_mixing} \cite{Aravinda_Sir_PRR}. Introducing the subset $w$ enables the said analytical calculation which subsequently gives the form in which equation Eq. \ref{max_mixing} is written. In practise, we initialise an ensemble of a 1000 local unitary equivalent dual unitary gates, independently sampling the local gates from the $w$ distribution, and then find the maximum mixing case. The value obtained for each case matches very well with the theoretical predictions made following Eq. \ref{max_mixing}, close to numerical precision. In order to source the actual dual unitary gates for the qubit case, we make use of the standard Cartan decomposition of the general $2-$qubit gate, specified as $U(\alpha, \beta, \gamma)$. This general Cartan gate follows the dual unitary conditions when $\alpha = \beta = \pi/4$, and the free parameter $\gamma$ specifies the entangling power of the gate which is given as \cite{Aravinda_Sir_PRR}:
	\begin{equation}
		e_P(U) = \frac{2}{3}cos^2(2 \gamma).
	\end{equation} 
	This allows us access to a set of gates satisfying dual unitarity with tunable entangling power, leading to a tunable ergodic nature \cite{, Dual_Algorithm}.
	
	Apart from dual unitary, we shall also consider the individual block $U$
	from the the class of solvable quantum circuits introduced in Ref. \cite{more_global_randomness}, which are also arranged in a brickwall configuration, along with local random unitaries $\{ u_i, v_i \}$. While the arrangement for this case is similar to the dual unitary construction that we have described above, the Floquet nature of the circuit is no longer retained, as the random unitaries are independently sampled at each space time point. Following from Ref. \cite{more_global_randomness}, it is known that a circuit constructed from the solvable gates shows a systematic convergence towards unitary $2$-designs upon varying the entangling power, while having a solvable form. Here, a unitary \(k\)-design denotes an ensemble whose statistical moments up to order \(k\) reproduce those of the global Haar random ensemble on \(U(2^N)\). A unitary \(2\)-design therefore reproduces the corresponding second order Haar moments and provides a useful measure of convergence towards Haar random dynamics.
	
	The underlying principal behind reservoir computing is mapping the input data into a higher dimensional space which then enables accurate predictions via classical postprocessing, usually involving a simple linear regression model. The ability of a quantum reservoir to map input data to this higher dimensional space upon being injected into the reservoir state, is quantified by the operator spread observed for the reservoir system in the Krylov space and correlates well with the long term saturation value of the Krylov complexity \cite{Wisniacki_Krylov_first, RabinKC1, RabinKC2}.
	
	Krylov space method have been developed for a variety of quantum systems, including those related to open system dynamics \cite{Krylov_open, Krylov_open_sahu}, and time dependent models \cite{Krylov_Time_dependent_Adolfo}, where they are used for characterising the dynamical behaviour \cite{Wisniacki_Chaos, Gill_Scrambling, asmi_halder_frozen_dynamics}. Most relevant for our present case is the discussion of Krylov space properties for the case of quantum circuits, which involves structured examples like the brickwall arrangement of gates and the trotterised implementation of unitary evolution generated by a many body Hamiltonian \cite{Claeys_Circuit_Krylov, Wisniacki_Krylov_Claeys}. 
	
	The class of dual unitary quantum circuits forms an example of maximally chaotic dynamics, leading to maximal operator spread in the Krylov space. However, it is interesting to note that even for the case of trotterised Hamiltonian evolution, the quantum circuit tends to have a behaviour similar to that of the maximally chaotic case of dual unitaries for larger trotter steps, irrespective of the underlying Hamiltonian being integrable and noninteracting \cite{Claeys_Circuit_Krylov, Wisniacki_Krylov_Claeys}. Thus, structured quantum circuits like those involving the brickwall geometry can be expected to have near maximal Krylov behaviour, similar to that of the dual unitary. This is due to the high trotter step of unit size that is built into the structure of the circuit.
	
	For a circuit model, an underlying Hamiltonian is not usually specified. Thus, the standard Liouville operation which is defined as the commutator of an initial operator with Hamiltonian is not directly applicable \cite{RabinKC1, RabinKC2}. Instead the initial operator $\hat{O}$, is specified under operator-state mapping, as a wavefunction in the Hilbert-Schmidt space and defined as $\ket{O}$. Under subsequent Heisenberg evolution and the Floquet nature of the circuit unitary $\mathcal{U}$, the evolved operator following $t$ steps is given as $\hat{O_t} = \mathcal{U}^{-t} \hat{O} \mathcal{U}^{t}$, with the corresponding mapped state being $\ket{O_t}$. The Krylov basis is obtained by orthogonalising the sequence of time evolved states given by $\{ \ket{O_1}, \ket{O_2} ...\}$ following a Gram-Schmidt process of obtaining orthonormal vectors from a linearly independent collection . The resulting orthonormal set of vectors is represented by $\{ \ket{\mathcal{O}_1}, \ket{\mathcal{O}_2} ...\}$ \cite{Claeys_Circuit_Krylov}. A basic outline of the approach with a working definition of Krylov complexity is provided in Appendix \ref{Krylov_basics}. 
	
	For time step $t$, the time evolved operator $\ket{O_t}$ maybe represented in terms of the orthonormal basis as the following superposition:
	\begin{equation}
		\ket{O_t} = \sum_{n = 0}^{t} \beta_{n,t} \ket{\mathcal{O}_n}.
	\end{equation} 
	from which the Krylov complexity is defined as:
	\begin{equation}
		\mathcal{K}_C^t = \sum_{n = 0}^{t} n |\beta_{n,t}|^2.
	\end{equation} 
	
	With a single site initial operator, under dual unitary dynamics, the analysis is very simplified. Owing to the maximally scrambling nature, the set $\{ \ket{O_t} \}$ is in itself orthonormal, and hence do not require the Gram-Schmidt procedure.  Thus, $\beta_{n,t}$ takes the value of unity when $t$ equals $n$ ($\beta_{n,t} = \delta_{n,t}$). This further implies that Krylov complexity $K_C^t$ has a maximal rate of increase. Further, it also implies that all the resulting Arnoldi coefficients also have the value of unity \cite{Claeys_Circuit_Krylov}.
	
	Given the basis of orthonormal vectors, we may define a unitary superoperator corresponding to the circuit operation $\mathcal{U}$ as a matrix upon the said basis. The elements of the matrix are given as $\mathcal{U}_{m,n} = \bra{\mathcal{O}_m}\mathcal{U}\ket{\mathcal{O}_n}$. The unitary superoperator can be characterised by the following coefficients, denoted by $a_n, b_n$ and $c_n$, which are defined as:
	\begin{align}
		a_n = \bra{\mathcal{O}_n}\mathcal{U}\ket{\mathcal{O}_n} \\
		b_n = \bra{\mathcal{O}_n}\mathcal{U}\ket{\mathcal{O}_{n-1}} \\
		c_n = \bra{\mathcal{O}_0}\mathcal{U}\ket{\mathcal{O}_n}
	\end{align}     
	Thus, for the case of dual unitary dynamics, $b_n = 1$, with $a_n, c_n = 0$. The unitary superoperator takes the form of a Hessenberg matrix, where all elements are zero, except of the principal off diagonal containing a series of $1$s \cite{Claeys_Circuit_Krylov}.
	\begin{figure}
		\includegraphics[width = \linewidth]{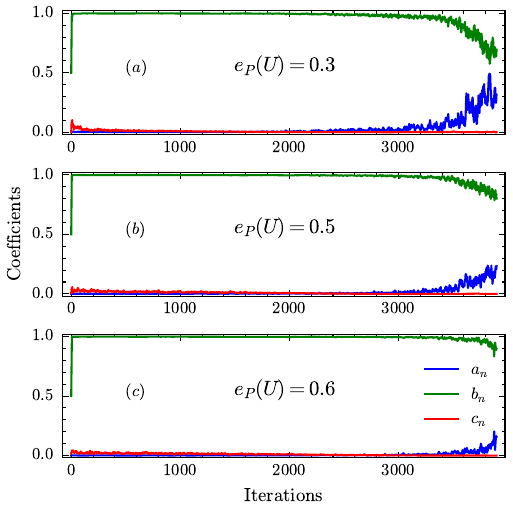}
		\caption{Variation of the coefficients $a_n, b_n$ and $c_n$ of the unitary superoperator $\mathcal{U}$ with increasing number of iterations for creating the orthonormal basis, under dual unitary dynamics. Note that upon decreasing the parameter of entangling power $e_P(U)$ causes an earlier departure from the theoretical behaviour found in the thermodynamic limit.}
		\label{Fig_deviation}
	\end{figure}
	
	The Krylov space behaviour specified above persists for arbitrarily large \(t\) in the thermodynamic limit, irrespective of the ergodic nature of the dual unitary circuit. However, for finite systems, deviations eventually arise because a restricted Hilbert space cannot lead to an indefinitely growing sequence of mutually orthogonal operators. These finite size effects also reveal the dependence of operator spreading towards circuit dynamics, as for weaker mixing, the initial operator spreads over a smaller set of orthogonal directions, causing an earlier departure from the ideal behaviour than in more strongly mixing circuits.
		
		For comparison, the conventional Liouvillian Krylov space has a maximal index \(K=d^2-d+1\), where \(d\) is the Hilbert space dimension. For \(N=6\), \(d=64\), $K$ is $4033$. Although this bound does not strictly apply to circuit dynamics, we use \(K\) as a reference scale for the number of iterations \cite{RabinKC2}, with each iteration corresponding to the application of single step circuit unitary. The finite system reproduces the expected thermodynamic limit behaviour over a substantial range of iterations, but deviations appear earlier for gates with smaller \(e_P(U)\), under the maximal mixing condition. In contrast, gates with larger entangling power and correspondingly stronger mixing retain the maximally scrambling behaviour for longer times. Such deviations from maximal Krylov growth provide a useful diagnostic of the underlying dynamics \cite{Wisniacki_Krylov_Claeys}, and their dependence on \(e_P(U)\) is shown in Fig. \ref{Fig_deviation}.
		
	\begin{figure}
		\includegraphics[width = \linewidth]{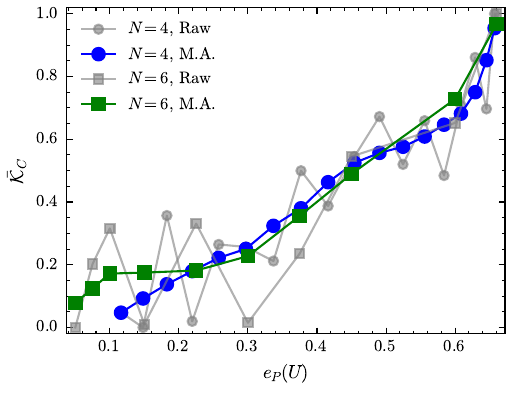}
		\caption{Variation of the Krylov complexity saturation with increasing value of entangling power $e_P(U)$ of constituent gates, for dual unitary circuits having system size $N = 4,6$. The increase in $\bar{\mathcal{K}}_C$ is indicative of improved task performance for circuits formed of gates with larger $e_P(U)$. The raw data points are shown in gray, with the coloured line portraying an approximate variation, representing the moving average (M.A.) obtained from the Savitsky-Golay filter. $\mathcal{K}_C$ values are normalised to enable comparison of saturation behaviour across different system sizes with $N$ being $4$ and $6$.}  
		\label{Figure_Krylov_Saturation}
	\end{figure}
	
	The finite system deviation also impacts the variation of Krylov complexity with time, where instead of increasing linearly, the variation with time ultimately saturates. The average value upon saturation $\bar{K}^{sat}_C$ again reflects the ergodic nature of the circuit and varies with changing the entangling power of the constituent gates, as shown in Fig. \ref{Figure_Krylov_Saturation}. This is an important observation which links the ergodic nature of the underlying circuit with Krylov complexity for the finite case. These said deviations are primarily a finite sized artifact of a limited Hilbert space. Other factors which influencing the deviation are numerical stability of Arnoldi iterations, which may be stabilised by orthogonalising with respect to all previous vectors, for the final few iterations where the deviations begin to emerge. We believe that this contributes to the fluctuations observed in Fig. \ref{Figure_Krylov_Saturation}, while the general trends is obtained by fitting an appropriate continuous graph through the points with a moving average obtained from the Savitsky-Golay filter. The variation of Krylov complexity is similar for \(N=4, 6\), indicating a weak system size dependence over the range considered, with \(\mathcal{K}_C\) normalized to \([0,1]\).
	
	Following from the earlier discussion, where average Krylov Complexity saturation $\bar{K}^{sat}_C$ was found as a reliable indicator of task performance in reservoir computing, the results presented above, provide us with an initial indication that increasing the ergodicity by appropriately changing the entangling power should lead to improved task performance. Atleast, with higher ergodicity, the input data is being mapped to a higher dimensional feature space. Within regular Hamiltonian dynamics the saturation value of Krylov complexity has a strong initial operator dependence, and is therefore a questionable indicator of quantum dynamics \cite{Aravinda_Sir_PRE, Wisnaicki_Saturation}. However, the impact of initial state dependence is somewhat reduced in the case of dual unitary dynamics as the initial growth rate is maximal, and the same for all considered examples of initial operators.

	\section{Synthetic Datasets and Associated Tasks for Benchmarking Performance}
	\label{datasets}
	
	\subsection{NARMA Task}
	
	For benchmarking the memory capability of the reservoir, we evaluate task performance on a nonlinear autoregressive moving average (NARMA) sequence. NARMA tasks are widely used as synthetic benchmarks in time series analysis and reservoir computing because of their inherent complexity and the nonlinear dependence of each input with the previous output values. Thus, the value which is to be predicted at time step $k+1$ depends nonlinearly on the output at step $k$, on a finite window of past outputs, and on delayed input values. Testing the reservoir for predictive task using the NARMA dataset tests two desirable properties of a dynamical learning system simultaneously, which is the fading memory capacity and the response to a complex, nonlinear mapping. 
	
	The window of previous outputs is determined by the order $M$ of the associated NARMA task. The general expression for a $M$th order NARMA is given as:
	\begin{equation}
		y_{k+1}
		=
		a\,y_k
		+
		b\,y_k\sum_{j=0}^{M-1} y_{k-j}
		+
		c\,s_{k-L+1}s_k
		+
		d,
	\end{equation}
	with the value of coefficients fixed to be:
	\begin{equation}
		(a,b,c,d)=(0.3,\,0.05,\,1.5,\,0.1).
	\end{equation}
	These are the standard NARMA coefficients used in Ref. \cite{Fuji_harnessing_disorder} along with other implementations related to time series forecasting. Unlike real datasets, the governing equation for the NARMA task is known exactly, and the difficulty of forecasting may be enhanced by increasing the order \(M\). Small orders probe modest memory requirements, while a larger order demand the retention and nonlinear combination of information over longer windows due to the structure of the NARMA expression \cite{Benchmarks}. This makes the benchmark particularly useful for assessing whether a reservoir merely reacts to the current input or genuinely builds an internal dynamical representation of the recent past. In the present work, the generated sequence of length $L$,
	\begin{equation}
		\{s_k,\; y_k\}_{k=0}^{L-1},
	\end{equation}
	serves as a controlled testbed for quantifying the temporal memory and nonlinear computational power of the reservoir. Thus, the forecasting task involves mapping an input datapoint $s_k$ to its probable output $y_k$, where the latter depends heavily on a finite window of previous outputs. 
	
	We shall make use of the NARMA task, for two different kinds of input data points $\{s_k\}$ in order to benchmark the performance of our circuit reservoir. For each case, the task performance would be evaluated for increasing values of the order $M$.
	
	Firstly, we take the input data points $\{s_k\}$ to be deterministic, and obtained as a product of three different sinusoidal functions with separate frequencies. Therefore we have:
	\begin{equation}
		s_k
		=
		0.1\left[
		\sin\!\left(\frac{2\pi \alpha k}{T_{\mathrm{f}}}\right)
		\sin\!\left(\frac{2\pi \beta k}{T_{\mathrm{f}}}\right)
		\sin\!\left(\frac{2\pi \gamma k}{T_{\mathrm{f}}}\right)
		+1
		\right]
	\end{equation}
	with $k=0,1,\dots,L-1$, where $L$ is the total length of the sequence. Here, we have:
	\begin{equation}
		(\alpha,\beta,\gamma)=(2.11,\,3.73,\,4.11) \,\,\,\,\
		\text{and} \,\,\,\,\
		T_{\mathrm{f}}=100.
	\end{equation}
	$T_f$ represents a large time interval over which the sinusoidal functions show independent variation owing to frequency parameters $(\alpha,\beta,\gamma)$.
		This specific form of the input is chosen due to its numerical stability while the additive constant and the prefactor \(0.1\) keep the input bounded in a small positive interval, thereby producing a smooth, weakly modulated signal suitable for controlled benchmarking.
	
	 In the second implementation of the NARMA task, the input data points $\{s_k\}$ are taken to be random. The other parameters, such as the coefficients $(a,b,c,d)$ remain unchanged. Here, each data point $s_k$ is independently sampled from a random uniform distribution between $0.0$ and $0.2$. Thus we have:
		\begin{equation}
			s_k \sim U(0.0,0.2) \,\,\,\,\ \text{for} \,\,\,\,\ k = 0,1,...,L-1
		\end{equation}
		
		Since the sinusoidal NARMA task uses a smooth deterministic input with a inherent temporal structure, a part of the information required for prediction is already encoded within the input sequence. It therefore probes the ability of a forecasting model to exploit both the intrinsic temporal regularity of the driving signal and the nonlinear memory dependence of the NARMA map. In contrast, for randomly sampled inputs the temporal correlations of the input signal are removed, providing a more stringent test of the model’s ability to reconstruct the required nonlinear temporal dependencies. For a reservoir implementation, this distinction is particularly relevant. With sinusoidal driving, temporal structure present in the input can assist the readout, whereas for random driving the required memory and nonlinear feature construction must arise predominantly from the internal reservoir dynamics. The random input NARMA task therefore provides a more direct probe of the reservoir’s intrinsic memory and nonlinear information processing capability.
	
	As we using both of the tasks to benchmark reservoir performance, we add a normalisation step for the generated dataset to make the performance metrics comparable across the two implementations. For this, apart from normalising the input $s_k$ between $0$ and $1$, required for the injecting process, we also normalise the output $y_k$ within the same limits.
	
	\subsection{Mackey-Glass Variation}
	
	As a benchmark for autonomous time series prediction, we consider the Mackey-Glass (MG) delay differential system, which is a standard synthetic dataset in reservoir computing and nonlinear time series analysis. This also follows from the discussion of Ref. \cite{Fuji_harnessing_disorder} where the MG task is introduced as a benchmark for learning autonomous dynamical systems, having chaotic features. A central feature of this system is the delay parameter \(\tau_{\mathrm{MG}}\), which controls the complexity of dynamics. For sufficiently small delay the dynamics remains regular with the autonomous prediction being trivial. For larger delay the variation becomes chaotic, with the crossover arriving at $\tau_{\mathrm{MG}} > 16.8$. In order to have those said chaotic features, we take $\tau_{\mathrm{MG}} = 17.0$ \cite{Fuji_harnessing_disorder}.
	
	In continuous time, the Mackey-Glass dynamics is governed by the following delayed nonlinear equation:
	\begin{equation}
		\frac{dx(t)}{dt}
		=
		\beta \frac{x(t-\tau_{MG})}{1 + x(t-\tau_{MG})^\omega}
		-
		\gamma x(t),
	\end{equation}
	where \(\beta\) controls the delayed nonlinear feedback, \(\gamma\) gives the linear damping, \(\omega\) is the nonlinearity exponent, and \(\tau_{MG}\) is the delay time. The nontrivial character of the dataset arises from the competition between dissipation and delayed feedback: the future state is not determined only by the present value \(x(t)\), but also by a past value \(x(t-\tau_{MG})\), so the system possesses an intrinsic memory. As \(\tau_{MG}\) is increased, this memory becomes sufficiently long to introduce complex behaviour in the trajectory $x(t)$ and generate chaotic behaviour.
	
	In our present implementation, the dataset is generated directly by
	the numerical solution of the Mackey-Glass equation. Writing the discrete sequence as \(x_t\), the update rule becomes:
	\begin{equation}
		x_{t+1}
		=
		x_t
		+
		dt\left[
		\beta \frac{x_{t-\tau_d}}{1+x_{t-\tau_d}^\omega}
		-
		\gamma x_t
		\right].
		\label{MG_working}
	\end{equation}
	Thus, unlike a memoryless synthetic signal, each new point is produced recursively from the current value and a delayed value from \(\tau_d\) steps in the past.
	
	The specific parameter choice for our present implementation is given as:
	\begin{equation}
		\beta = 0.2,\qquad
		\gamma = 0.1,\qquad
		\omega = 10,\qquad
		dt = 0.1,
	\end{equation}
	with $\tau_d = \tau_{MG}/dt$ taken as a control parameter having the value $170$. Following from this numerical setup the datapoints $x_t$ are separated from each other by the time interval $dt$. Instead of solving the numerical equation explicitly, the MG time series can also be obtained using standard python libraries such as ReservoirPy \cite{ReservoirPy}, after specifying the above parameters. It is important to subsample the obtained numerical solution to ensure that the data points generated are separated by a unit interval of time.
		
		Thus, from the differential equation we have the numerical solution $\{x_t\}$, where each element is separated from the next by a unit time interval. For prediction purposes, the resulting time series is typically organized into a few step ahead input and target pairs. This window is defined by the symbol $\alpha$. If the generated sequence is denoted by \(\{x_t\}\), then the learning task is to infer the map
		\begin{equation}
			x_t \longmapsto x_{t+\alpha},
		\end{equation}
		or more generally, to use the available state at time \(t\) to predict the future point in the trajectory, $\alpha$ steps ahead. Following our convention, the input-output pair $\{ s_k, y_k\}$ is therefore $\{x_t, x_{t+\alpha}\}$. For the chaotic regime this prediction is a nontrivial test because nearby trajectories separate exponentially fast and prediction over larger values of $\alpha$ tend to saturate.
	
	The above map is an example of a teacher-forced prediction, where predicting each output point $x_{t + \alpha}$ requires the explicit input of the point $x_t$. A more stringent test is the prediction of a chaotic time over a sufficiently longer horizon, without the explicit input of nearby preceding data points. For this autonomous prediction, the reservoir is first trained in a teacher-forced manner using \(10000\) consecutive time steps, where the true MG value at each step is supplied as the input and the subsequent value is used as the target. After training, the feedback loop is closed and the predicted output is recursively supplied as the input for the next reservoir step. The resulting autonomous trajectory is then evolved for a further \(2000\) time steps and compared with the corresponding MG reference sequence, both through finite horizon prediction errors and also the reconstructed time series attractor. Here, small prediction error accumulates rapidly over time, owing to the chaotic nature of the underlying MG time series, although the overall attractor of the generated sequence matches the original. 
		
		Consequently, short time prediction requires the model to capture the local nonlinear flow accurately, while long horizon prediction, although fundamentally limited by the sensitive dependence on initial conditions, forms a stringent test of true autonomous predictions from the reservoir. For this reason, the MG dataset is regarded as a standard benchmark in time series analysis. It is simple enough to be generated from a known equation, yet sufficiently rich to probe key capabilities of a learning system involving nonlinear modelling, delayed memory, autonomous evolution, and robustness in the presence of chaos. In the present work, it therefore serves as a controlled synthetic dataset for testing whether the model can learn and predict the evolution of a nonlinear delayed dynamical system.
	
	\section{Characterising Reservoir Performance for Dual Unitary Setup}
	\label{Benchmarking}
	
	We shall now present our results for task performance attained by the dual unitary circuit based reservoir, for different tasks based on the synthetic datasets that we have defined in the previous section. In all of these results, the multiplexing parameter is $V = 6$, unless specified, and is same as the number of qubits. Doing so allows us to access task performance in the most efficient way. Later, in the following section, we shall also discuss about the role of multiplexing and explain the reason behind taking $V = N$ as the default value of our multiplexing parameter. Before discussing task performance obtained by the specific reservoir, it is necessary to evaluate the actual contribution made by the reservoir towards a predictive task, beyond performance attained by the final classical post processing step acting on its own, without the reservoir. In this setup, the role of a reservoir is to generate a higher dimensional feature space which enhances the predictions made by the final classical postprocessing step. Since in our case, the final step is the simple linear regression model, we shall compare results of the reservoir against linear regression acting by itself on the same task.
	
	We start with analysing the memory capacity of the dual unitary circuit functioning as the reservoir. For this, we use the NARMA dataset of increasing orders, and with the task being to accurately predict the mapping between a given $s_k$ and its corresponding $y_k$. This task is evaluated for increasing values of $e_P(U)$, which we know from our previous discussion, increases the mixing rate of the associated dynamics. As mentioned before, the local random unitaries are chosen such that the mixing rate is maximised for the particular value of $e_P(U)$ that is being considered. Here, we also introduce two specific version of linear regression models to compare the reservoir performance against. These are described as follows:
		\begin{itemize}
			\item Single input linear regression: Given the pair $(s_k, y_k)$, we optimise over parameters $w_0$ and $w_1$, where $y_k = w_1 s_k + w_0$. 
			
			\item $NV$ delayed input linear regression: For a reservoir with \(N\) qubits and multiplexing \(V\) times, we introduced the classical regression model containing \(NV\) delayed input values. For this case, we have: $y_k = w_0 +\sum_{j=0}^{NV-1}w_{j+1}s_{k-j}$, with $\{w_0, w_1 ... w_{NV}\}$ being the parameters that are optimised using linear regression.
		\end{itemize}
		\begin{figure}
			\includegraphics[width = \linewidth]{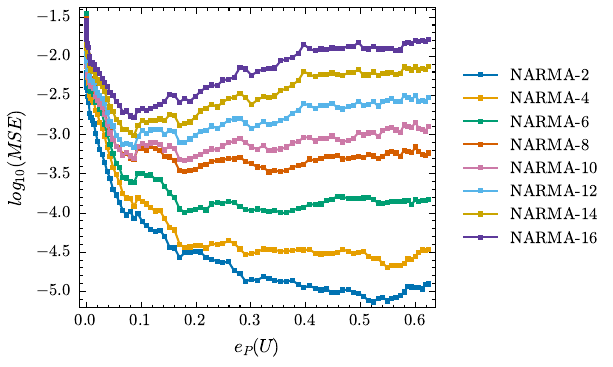}
			\caption{Variation of task peformance for dual unitary reservoir, with respect to entangling power $e_P(U)$ of the constituent gates for increasing orders of NARMA tasks. Here, each gate is chosen to maximise the associated mixing rate. The input for each NARMA task is sinusoidal in nature. Here $N = 6$ and also $V = 6$. The presented results are smoothened using the Savitsky-Golay filter to suppress fluctuations between neighbouring points.}  
			\label{Figure_Narma_vs_ep}
		\end{figure}

		The optimised weights are obtained from the training data using the Moore-Penrose pseudoinverse, and the trained model is subsequently applied to the held out evaluation data. The resulting prediction error is quantified using \(\log_{10}(\mathrm{MSE})\). The reservoir makes no appearance in these calculations.
		
		In the first case, only the instantaneous input \(s_k\) is provided to the regression model, and hence represents the 'no-reservoir' linear baseline. The same benchmark has also been considered in Ref. \cite{Fuji_harnessing_disorder}. Compared to that, in the second case, the regression model is supplied explicitly with an \(NV\)-step temporal history of the input. This represents a more stringent test for the reservoir, and constitutes an 'explicit-memory' baseline. Thus, when a reservoir outperforms the latter linear regression baseline, it indicates that the features reconstructed via multiplexing are superior than the same number of actual data points being explicitly provided to a linear regression model. 
	\begin{figure*}
		\includegraphics[width = \linewidth]{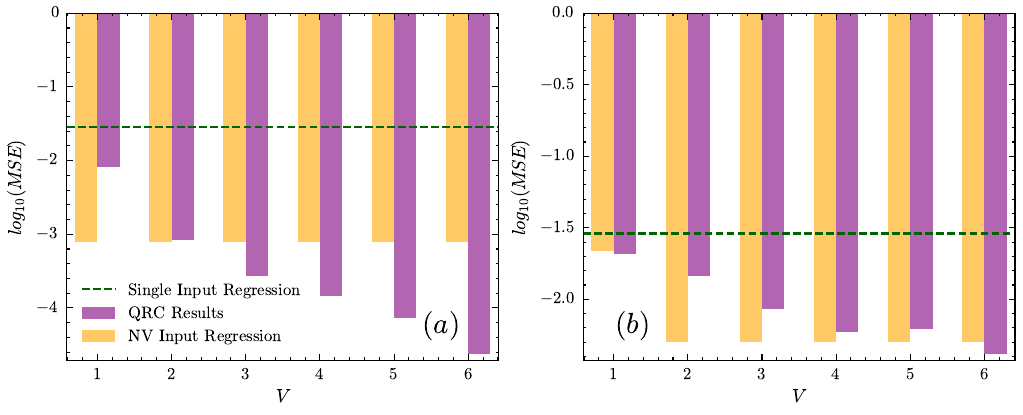}
		\caption{A comparison of prediction accuracy for the dual unitary reservoir against the single input and the $NV$ delayed input linear regression for different values of multiplexing parameter $V$. Here, the value of $e_P(U) = 0.3$ for the gates involved in the dual unitary circuit. Task performance is evaluated on NARMA order $6$ for $(a)$ Sinusoidal input and $(b)$ Random inputs.}
		\label{Figure_QRC_vs_LR_Narma}
	\end{figure*}
	
	In Fig. \ref{Figure_Narma_vs_ep} we have shown the variation of task performance for the dual unitary reservoir for increasing values of $e_P(U)$. This task performance is evaluated for the sinusoidal input NARMA dataset of increasing orders. Here, it is clear that NARMA tasks with higher orders perform worse than examples having lower orders. This is because of the fading memory capacity associated with the scrambling behaviour of the underlying circuit, which prevents retention of features required to accurately predict a higher order NARMA series.
	
	Another observation is related to the fact that for lower order NARMA tasks, the performance of reservoir increases nearly monotonically, whereas for higher order NARMA tasks, there is a pronounced dip indicating maximal performance. Beyond this dip, the prediction accuracy worsens and then plateaus at its respective saturation values. These are also systematically arranged according to the associated NARMA order.
	
	A qualitatively similar result is also obtained for the random input NARMA task, however the task performance is on average worse off by two orders of magnitude. With this result, we pick out a moderate NARMA order ($M = 6$), along with a moderate value of $e_P(U)$, being $0.3$ and compare the performance obtained from the reservoir against linear regression benchmarks. This comparison is carried out for both the sinusoidal and random input NARMA tasks. For each value of $V$ we take $NV$ previous inputs in the delayed input linear regression. The comparison is presented in Fig. \ref{Figure_QRC_vs_LR_Narma}. Here, the quantum reservoir outperforms the single input linear regression in all cases. Further, it also outperforms the $NV$ delayed input linear regression for moderately high values of $V$ in the case of a sinusoidal input NARMA task. With increasing $V$ the difference in performance widens in favour of the reservoir. In case of the random input NARMA task, the performance of reservoir nearly matches that of the $NV$ delayed input linear regression. 
	
	These results indicate that the improvement over single-input regression originates from the additional temporal and nonlinear features generated by the reservoir dynamics. The \(NV\)-delay regression provides a substantially more stringent benchmark, since the relevant input history is supplied explicitly rather than being retained and processed dynamically by the reservoir. For the sinusoidal input, increasing multiplexing allows the reservoir to construct a progressively richer feature representation and eventually outperform this delayed-input model. In contrast, for the temporally uncorrelated random input, the two approaches attain comparable performance, indicating that the advantage of the reservoir over an explicit-memory linear model is conditional on the structure of the input and the available multiplexing depth.
	
	We shall now present results for the teacher-enforced and autonomous predictions of the MG task. While the NARMA task is an important probe for investigating the fading memory of the reservoir, prediction involving the MG task studies the actual ability of the reservoir towards reconstruction and prediction of an autonomous nonlinear dynamical system. Studying the task performance gives evidence of improved prediction accuracy using the circuit reservoir in 'edge-of-chaos' regime, as found in other examples of quantum reservoirs \cite{kobayashi_edgeofchaos}. A similar feature was also observed in the previously discussed NARMA tasks, where for higher orders there was a noticeable dip in task prediction errors at relatively small values of $e_P(U)$. Increasing the entangling power further, lead to a decrease in prediction accuracy.
	\begin{figure*}
		\includegraphics[width = \linewidth]{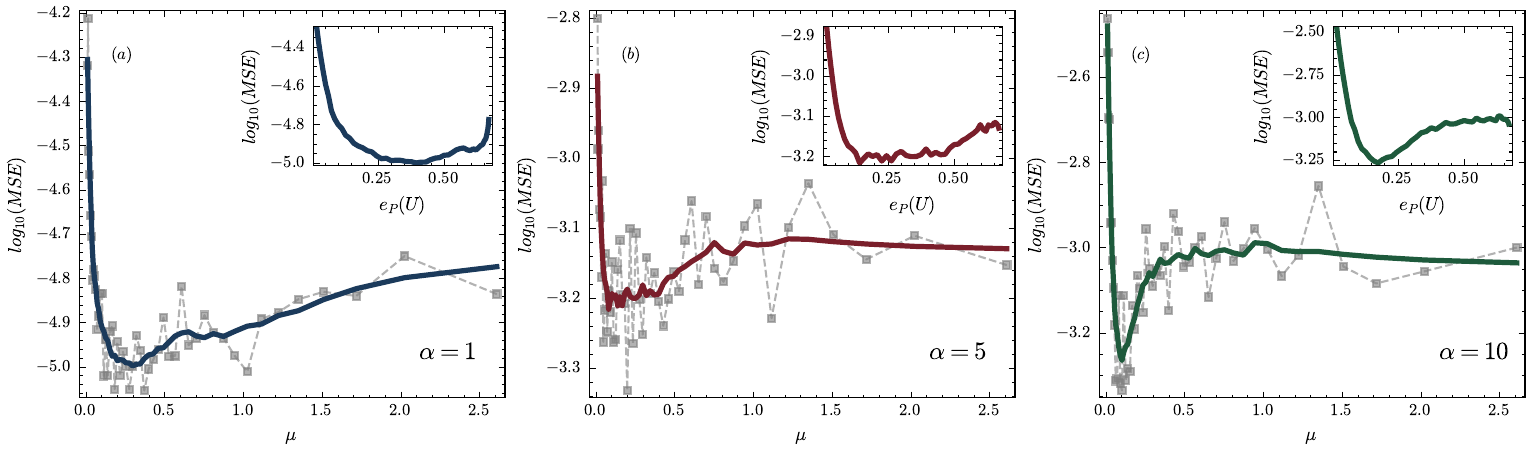}
		\caption{Variation of task performance for Mackey-Glass time series with the mixing rate of the overall circuit. Here the task performance corresponds to optimising parameters for the map $x_t \rightarrow x_{t+\alpha}$ for $(a)$ $\alpha = 1$, $(b)$ $\alpha = 5$ and $(c)$ $\alpha = 10$. The same variation is show in inset with respect to the entangling power $e_P(U)$ of the associated gates. The dotted lines show the variation of ensemble average over a few realisations of reservoir unitary with different local random unitaries. The colored line indicating the moving average obtained from the Savitksy-Golay filter.}
		\label{Figure_MSE_vs_mixing_MG}
	\end{figure*}
	
	The same features are also found when analysing the predictive capacity of the reservoir in context of the Mackey-Glass task. Here, the objective is to learn the mapping $x_t \rightarrow x_{t + \alpha}$ for datapoints $x_t$, which are solutions to the MG differential equations, subsampled to have a unit time step separation between them. We take $\alpha = 1,5 \,\, \& \,\, 10$. In order to better highlight the role of underlying dynamics towards task prediction, we plot the variation of performance with the mixing rate $\mu$ of the global circuit evolution. For the class of dual unitary circuits, $\mu$ is directly related to the entangling power of gates, and is given by $\mu = -\ln(1 - 3* e_P(U)/2)/3$.
	
	Observing the variation of task performance in Fig. \ref{Figure_MSE_vs_mixing_MG} the best prediction accuracy is obtained close to a non zero value of mixing, and it then reduces gradually with increased mixing. For $\alpha = 5$ and $\alpha = 10$ the gradual reduction saturates into a fixed value. Further, note that the prediction accuracy for $\alpha = 5$ and $\alpha = 10$, while significantly lower than the case of $\alpha = 1$, are relatively similar to each other. The variation of task performance with the tunable entangling power parameter is shown inset. Here, we also observed significant fluctuations within neighbouring points. In order to stabilise the variation we carried out an ensemble average over few (10, in our case) reservoir realisations, with different local random unitaries. In each case the mixing rate was verified to be within $10^{-3}$ of the maximal value for the given entangling power. The variation between ensemble members were actually not too different, and in the end upon averaging, we still got some persistent fluctuations. In the displayed result we have also carried out a moving average to stabilise the result over neighbouring fluctuations, and is obtained using the Savitsky-Golay filter.  Here we also mention that the reservoir significantly outperforms the no-reservoir baseline for linear regression. The observed values of $log_{10}(MSE)$ were $-2.89, -1.58, \,\, \& -1.21$ respectively for $\alpha = 1,5 \,\, \& \,\, 10$.
	\begin{figure}
		\includegraphics[width = \linewidth]{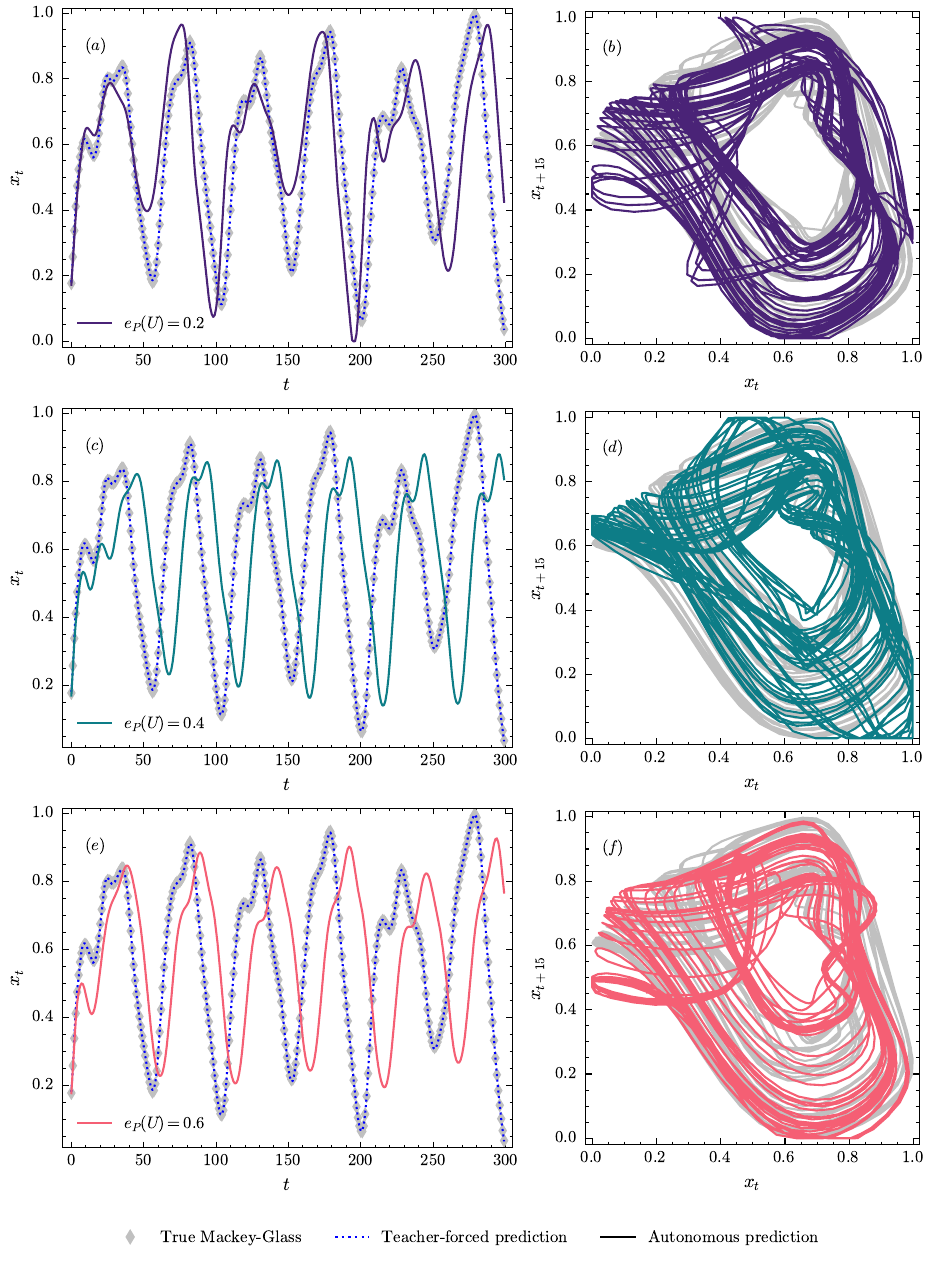}
		\caption{Variation of autonomous task prediction for the Mackey-Glass time series, in relation to the actual variation and the teacher-forced predictions. We show three specific values of $e_P(U)$ being $(a)$ $0.2$, $(c)$ $0.4$ and $(e)$ $0.6$. While the autonomous predictions show a deviation from the actual, the resulting attractor very nearly matches the  true variation in each case, as shown in $(b), (d) \,\, \& (f)$} 
		\label{Figure_MG_Autonomous}
	\end{figure}
	
	We shall now present results for the autonomous task prediction, which access the ability of reservoir for long-horizon forecasting. As described in a previous section, the task involves training the model over $10000$ steps in the regular teacher-forced prediction, following which the next $2000$ steps are predicted using the output generated by the reservoir. In this case, particularly for the Mackey-Glass task which has a chaotic behaviour, the small errors arising out of the initial autonomous predictions buildup overtime. This introduces a fundamental limit upon the accuracy that can be attained in an autonomous task prediction. Comparing the time series generated via autonomous and teacher-forced predictions in Fig. \ref{Figure_MG_Autonomous} it is apparent that while the latter tracks almost perfectly with the actual variation, there is a noticeable deviation in the former, after a small window of predictions. 
		
\begin{figure}
		\includegraphics[width = \linewidth]{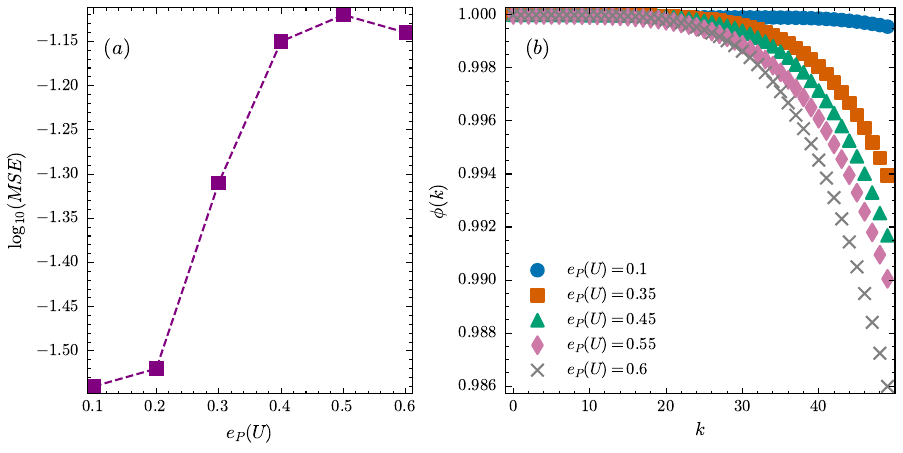}
		\caption{Measures for memory retention in reservoir dynamics for $(a)$ Autonomous predictions accuracy, upon ensemble averaging, for the first 200 steps with respect to entangling power $e_P(U)$ and $(b)$ Memory capacity $\phi(k)$ as a function of $k$ with $e_P(U)$ appearing as a parameter.}  
		\label{Figure_MG_Memory_autonomous}	
	\end{figure}

		Although the autonomous prediction does not track exactly, nevertheless the attractor resulting from the autonomous prediction does resemble the same obtained form the actual MG time series. Hence, the generated result continues to have the chaotic nature as the time series it is required to forecast. In order to quantitatively track prediction accuracy we measure the MSE value of the autonomous prediction with respect to the true variation over the first $200$ steps within the autonomously generated datapoints. The result are shown in Fig. \ref{Figure_MG_Memory_autonomous}. For each instance of entangling power we carry our an ensemble average over 10 reservoir implementations, as described above for Fig. \ref{Figure_MSE_vs_mixing_MG}. Under repeated implementations, the differences in MSE values between instances is small. The variation shown in Fig. \ref{Figure_MG_Autonomous} is one specific implementation from the few carried out. In Fig. \ref{Figure_MG_Memory_autonomous}, it is interesting to note that with higher values of $e_P(U)$ the autonomous predictions are increasingly worse off. This indicates that the memory retention of the reservoir diminishes under enhanced ergodicity of the associated dual unitary dynamics. 	 
	
	We consider another statistical parameter that in reference to retention of information, called as the memory capacity. Although defined for the conventional echo state network \cite{Stepney2024, Benchmarks} with for a random input signal, the quantity is independent of the underlying reservoir and instead depends only on the $Z$ matrix that is used for making predictions following the process of linear regression. In our work, we define the parameter as $\phi(k)$, which denotes the ability of a reservoir to construct the time series that is lagging by $k$ steps from the entries present in the $Z$ matrix. Instead of random inputs, we use as input the MG time series data on which task performance in analysed.
	
	The quantity measured by $\phi(k)$ can be understood as follows: Given the $Z$ matrix having the order $NV \times L$ we discard the $k$ initial components giving us the matrix $Z^*$ having order $NV \times (L-k)$. Using the matrix $Z^*$ we attempt to reconstruct the time series $u^* = \{ y_j \}_{j = 1}^{L - k}$. Following the same train/test ratio we have sequences $u^*_{\mathrm{test}}$ and $u^*_{\mathrm{train}}$ with which  we obtain the optimal parameters $w^*$. We therefore have:
	\begin{align}
		w^* = {Z^{*}}^{+}_{\mathrm{train}} \cdot u^*_{\mathrm{train}} \,\,\,\,\ \& \\
		\bar{u}_{\mathrm{pred}} = {Z^{*}}^{T}_{\mathrm{eval}} \cdot w^*.
	\end{align} 
	Here, $\bar{u}_{\mathrm{pred}}$ denotes the reconstructed sequence that lags by $k$ steps from the inputs present in the matrix $Z^*$. $Z^*_{\mathrm{train}}$ and $Z^*_{\mathrm{test}}$ are obtained by splitting $Z^*$ according to the training and evalutaion split used in the partition for $u^*_{\mathrm{test}}$ and $u^*_{\mathrm{train}}$. With access to the sequences $\bar{u}_{\mathrm{pred}}$ and $u^*_{\mathrm{test}}$, we can now evaluate the memory capacity $\phi(k)$ using the following expression:
	\begin{equation}
		\phi(k) = \frac{\mathrm{Cov}(\bar{u}_{\mathrm{pred}}, u^*_{\mathrm{test}})^2}{\mathrm{Var}(\bar{u}_{\mathrm{pred}}) \times \mathrm{Var}(u^*_{\mathrm{test}})}
	\end{equation}
	
	The results for memory capacity are shown in Fig. \ref{Figure_MG_Memory_autonomous}, which reflect the observations made from the autonomous prediction accuracy. Increasing the value of $e_P(U)$ and hence the mixing rate associated with the dynamics, memory capacity measure shows a systematic suppression, and is indicative of diminishing memory retention. These aspect indicate a decrease in task performance with increasing the nature of ergodicity associated with the reservoir dynamics.
	
	We know from Fig. \ref{Figure_Krylov_Saturation} that while increasing the ergodic nature of the reservoir unitary leads to a higher dimensional feature space mapping of the input data. This is useful for having an accurate task prediction. One the other hand, enhancing the mixing rate is also accompanied by a decrease in memory retention. This is detrimental towards task performance. Since an optimal reservoir balances the features of high dimensional mapping with memory retention, the edge-of-chaos becomes a viable regime for reservoir operation following from these heuristic arguments.
	
	\section{Role of Multiplexing}
	\label{Multiplexing}
	
	The role of multiplexing, as defined in Section \ref{Brickwall_Reservoirs}, is significant in the reservoir computing paradigm and acts as a feature generator leading to $NV$ nodes for each input datapoint. Thus, increasing $V$ leads to greater performance towards task prediction as it provides more nodes for the final classical postprocessing step to optimise against, thereby generating a more effective mapping between the input and output pairs.
	
	In this regard, the Krylov space analysis again becomes useful by estimating the optimal levels of multiplexing. As the key function of the reservoir unitary is to map an input into a higher dimensional feature space, the same can be used to evaluate till what point repeated multiplexing is useful. The states that originate from the application of reservoir unitary together generate the higher dimensional space which is a linearised span of all the generated state vectors \cite{Wisniacki_Krylov_first}. Thus, the state acted upon by the reservoir unitary several times only contributes to the feature space till the number of iterations where it remains independent to all the previously evolved state vectors. 
	
	A quantitative analysis of the above intuition has been shown to be possible using Krylov vectors generated from an initial set of operators, which are taken as the observables for which the expectation value is evaluated \cite{Cindrak_Krylov_Expressivity, Cindrak_Krylov_Observability}. Here, calculating the fidelity between time evolved states, once verified to be linearly independent, defines the measure of Krylov Observability which can be used to capture the feature space dimensions. This measure increases and subsequently saturates with increasing multiplexing, following the expected increase in fidelity between the involved states as the Krylov space saturates. The subsequent saturation of Krylov Observability also matches the onset of maximal task performance, which does not substantially improve with further multiplexing. This indicates the saturation of phase space formed by time evolved states, and is verified for the case of Hamiltonian evolution involving the observables $Z_i$ \cite{Cindrak_Krylov_Expressivity, Cindrak_Krylov_Observability}.
	
	Although native to a quantum circuit picture, the sequence of linearly independent time evolved states constructing the Krylov basis remains valid even for a continuous time Hamiltonian model \cite{Cindrak_Krylov_Observability}. 
	
	
	Consider an initial observable $O$, which is evolved under Heisenberg evolution:
	\begin{equation}
		O(t)=e^{iHt} O e^{-iHt}.
	\end{equation}
	The space generated by the time evolved operators $\{O(t_1),O(t_2),\ldots\}$ constitute a viable Krylov space \cite{Cindrak_Krylov_Observability}. Now, for a collection of observables $\{O_i\}_{i = 1}^{N}$ sampled at equally spaced discrete times $\{ t_k \}$, separated by an interval $\tau$, it is possible to construct the corresponding linearly independent space $\mathcal{F}^{(i)}$. Defining $\tau = T/V$ for a total time $T$, the Krylov observability of the $i$th observable is then introduced as
	\begin{equation}
		p_i(T) = \sum_{k = 1}^{R_i} \Big( 1 - F(O_i(t_k),O_i(t_{t_{k-1}})) \Big),
	\end{equation}
	where $F$ denotes the normalized fidelity between two time evolved operators and $R_i = \text{min}(V,dim(\mathcal{F}^{(i)}))$. The total Krylov observability for multiplexed measurements is obtained by summing over all measured observables, $\mathcal{O}(T)=\sum_{i = 1}^{N} p_i(T)$. Thus, unlike standard Krylov complexity, which characterizes spreading within a fixed Krylov basis, Krylov observability quantifies how many linearly independent operator directions are effectively accessed by the measured dynamics, and serves as a proxy for the accessible phase space dimension of the reservoir \cite{Cindrak_Krylov_Observability}. Finally, it has been found that the Krylov Observability shares a strong statistical correlation with Information Processing Capacity (IPC), which is a standard measure for evaluating reservoir performance \cite{Cindrak_Krylov_Observability, Soriano_IPC}.
	\begin{figure}
		\centering
		\usetikzlibrary {decorations.pathmorphing, decorations.pathreplacing, decorations.shapes }
		\resizebox{\columnwidth}{!}{%
			
			\begin{tikzpicture}
				
				\draw[thick] (2.0,2.0) -- (18.0,18.0);
				\draw[thick] (0.0,4.0) -- (14.0,18.0);
				\draw[thick] (0.0,8.0) -- (10.0,18.0);
				\draw[thick] (0.0,12.0) -- (6.0,18.0);
				\draw[thick] (0.0,16.0) -- (2.0,18.0);
				
				\draw[thick] (4.0,2.0) -- (0.0,6.0);
				\draw[thick] (6.0,4.0) -- (0.0,10.0);
				\draw[thick] (8.0,6.0) -- (0.0,14.0);
				\draw[thick] (10.0,8.0) -- (0.0,18.0);
				\draw[thick] (12.0,10.0) -- (4.0,18.0);
				\draw[thick] (14.0,12.0) -- (8.0,18.0);
				\draw[thick] (16.0,14.0) -- (12.0,18.0);
				\draw[thick] (18.0,16.0) -- (16.0,18.0);
				\draw[thick, dotted] (18.0,16.0) -- (-1.0,16.0);
				\draw[thick, dotted] (-1.0,3.0) -- (-1.0,16.0);
				
				\draw[thick, dotted] (4.0,2.0) -- (4.0,0.0);
				\draw[thick, dotted] (6.0,4.0) -- (6.0,0.0);
				\draw[thick, dotted] (8.0,6.0) -- (8.0,0.0);
				\draw[thick, dotted] (10.0,8.0) -- (10.0,0.0);
				\draw[thick, dotted] (12.0,10.0) -- (12.0,0.0);
				\draw[thick, dotted] (14.0,12.0) -- (14.0,0.0);
				\draw[thick, dotted] (16.0,14.0) -- (16.0,0.0);
				
				\foreach \i in {0,...,7}
				{
					\draw [thick,fill=purple,rounded corners=2.2pt] (2.5+2*\i,2.5+2*\i) rectangle (3.5+2*\i,3.5+2*\i);	
				}
				
				\foreach \i in {0,...,6}
				{
					\draw [thick,fill=purple,rounded corners=2.2pt] (0.5+2*\i,4.5+2*\i) rectangle (1.5+2*\i,5.5+2*\i);
				}
				
				\foreach \i in {0,...,4}
				{
					\draw [thick,fill=purple,rounded corners=2.2pt] (0.5+2*\i,8.5+2*\i) rectangle (1.5+2*\i,9.5+2*\i);
				}
				
				\foreach \i in {0,...,2}
				{
					\draw [thick,fill=purple,rounded corners=2.2pt] (0.5+2*\i,12.5+2*\i) rectangle (1.5+2*\i,13.5+2*\i);
				}
				\draw [thick,fill=purple,rounded corners=2.2pt] (0.5,16.5) rectangle (1.5,17.5);
				
				\draw[decorate,decoration={coil,aspect=0,segment length=12pt}, color=blue] (-1,1) -- (16,1);
				
				
				\draw[thick] (1,1) -- (2.0, 2.0);
				\draw[thick] (1,1) -- (-1.0, 3.0);
				\pgfmathsetseed{27}
				
				\foreach \i in {0,...,3}
				{
					\draw[thick, fill=pink] (1+4*\i,1) circle (0.5);
					
					\pgfmathsetmacro{\randAngle}{rnd*75}
					\pgfmathsetmacro{\randLength}{rnd*2 + 1}
					
					\draw[->, line width=1.5pt, red]	
					(1+4*\i,1) -- ++(\randAngle + 30:1.2);
					\draw[-, line width=1.5pt, red]
					(1+4*\i,1) -- ++(180 + \randAngle + 30:1.2);
				}
				
				\node[font = \Huge] at (3,-1) {Input: $s_k$};
				\node[font = \Huge] at (-1.25,-1.25) {$\sigma_{1}^z(t)$};
				
				\draw[->, >=latex, line width=5pt] (1.5,0.5) -- (2.5,-0.50);
				\draw[->, >=latex, line width=5pt] (0.5,0.5) -- (-0.5, -0.5);
				
			\end{tikzpicture}

		}
		\caption{A schematic illustration showing the action of a time evolved measurement operator $\sigma^z_i$ after $t = N$ time steps. $N = 4$ is the number of qubits, and the figure shows that after $N$ steps of time evolution the input $s_k$ spreads throughout the finite system, under periodic boundary conditions.}
		\label{Figure_ckt_diagram}
	\end{figure}
	
	Calculating Krylov Observability for the case involving circuit unitary leads to a different picture, wherein the saturation happens over several steps. For example, in case of maximal scrambling under dual unitary dynamics the unitary time evolution leads to orthonormal states, have vanishing fidelity with respect to each other. As discussed in Section \ref{Duals_section}, this is a finite size effect, and is governed by the ergodicity associated with the global unitary. This observation is also important in context of the trotterised Hamiltonian evolution as well, for it has been shown that under large trotter time steps the overall time evolution converges to the maximal scrambling case, for cases of both integrable and nonintegrable models \cite{Claeys_Circuit_Krylov, Wisniacki_Krylov_Claeys}. Thus, even Hamiltonian evolution under high trotter steps would lead to a similar behaviour for the Krylov Observability metric.   
	\begin{figure}
		\includegraphics[width = \linewidth]{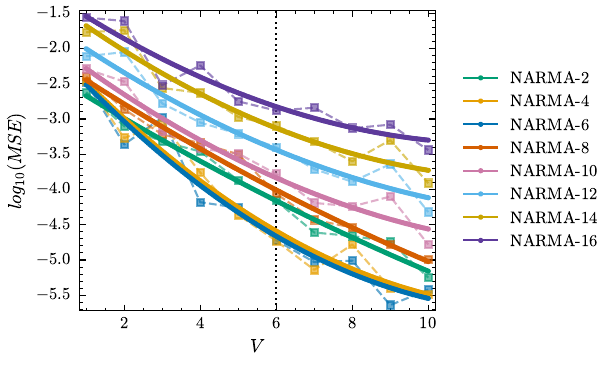}
		\caption{Variation of task performance with respect to the multiplexing parameter $V$ for increasing NARMA orders. $e_P(U) = 0.3$ for the gates used in constructing the dual unitary reservoir. The solid points are the actual datapoints whereas the solid line indicates the points obtained for fitting a quadratic order polynomial.}
		\label{Figure_multiplex_Narma}
	\end{figure}
	
	Following the above discussion would indicate that for circuit reservoirs increased multiplexing would always lead to direct improvements in the associated feature space, leading to better prediction accuracy. This increase, while also being aided by the classical postprocessing step, would primarily be driven by the underlying dynamics generated by the reservoir. The effect of latter, however has a finite capacity of influencing the task performance as we shall now discuss. This is also observed in practise where we find that the task performance saturates at multiplexing around $~N$, and a further increase has diminishing returns towards task performance. An explanation of this, for the dual unitary case is via the light cone argument. 
	
	By the geometric construction, in Fig. \ref{Figure_ckt_diagram} the time evolved measurement operator $Z_i$, after $N$ steps of time evolution is spread upon the entire wavefunction of the $N$ qubit state. We refer to this collection of $N$ time evolved states as 'intermediate'. This implies that the input data mapped onto the first qubit of the wavefunction is spread throughout the system and is reflected in the collection of 'intermediate' states. Further action of unitary evolution generates a state which is still orthonormal to the intermediate states, yet is somewhat redundant in terms of the information that is extracted by subsequent single qubit measurements. 
	\begin{figure}
		\includegraphics[width = \linewidth]{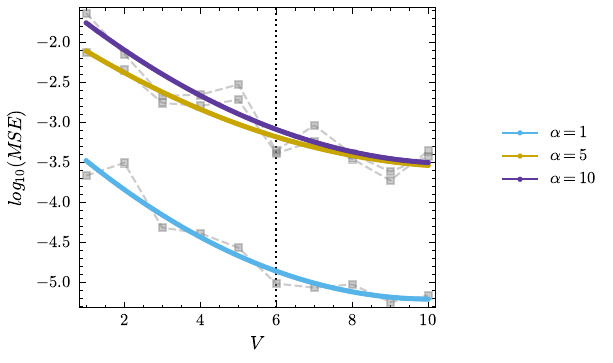}
		\caption{Variation of task performance with respect to the multiplexing parameter $V$ for the Mackey-Glass task for separate prediction windows $\alpha$. $e_P(U) = 0.3$ for the gates used in constructing the dual unitary reservoir. The solid points are the actual datapoints whereas the solid line indicates the points obtained for fitting a quadratic order polynomial.}
		\label{Figure_multiplex_MG}
	\end{figure}
	
	The variation of task performance, for a fixed value of ergodic mixing, and with respect to increasing orders of multiplexing is shown in Figs \ref{Figure_multiplex_Narma} and \ref{Figure_multiplex_MG} for the sinusoidal input NARMA and Mackey-Glass tasks respectively. While the individual points tend to converge, a more clear picture is obtained by fitting a simple polynomial curve with the given points. From this, the saturation of task performance with increasing $V$ becomes apparent, across a range of memory requirements.
	
	There is another reason for the saturation of task performance with increased multiplexing, and that is related to the convergence of reservoir state towards a featureless thermal state. Under maximal scrambling dynamics generated via dual unitary circuits, $V$ steps of evolution implies that the state is close to the thermal state, and hence subsequent measurements give very little further advantage for the reservoir computing task of time series prediction from the dynamical context. 
	
\begin{figure}
		\includegraphics[width = \linewidth]{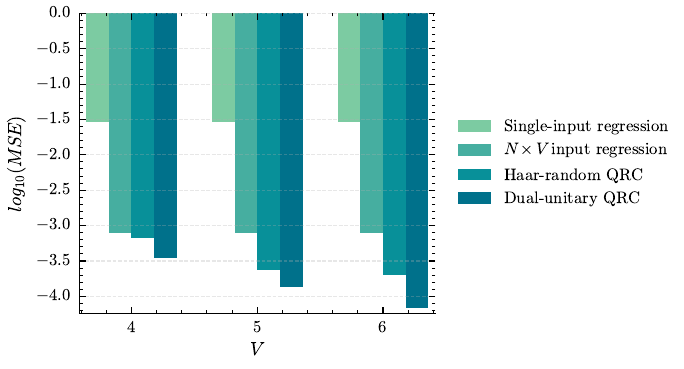}
		\caption{A comparision between dual unitary reservoirs and the same brickwall circuit made up of 2 qubit Haar random unitaries. The single input and $NV$ delayed input classical regression benchmarks are also provided for reference.}
		\label{Figure_haar_vs_dual}
	\end{figure}

	To demonstrate that the convergence towards a Haar random like featureless state reduces the task performance, we have compared the dual unitary reservoir against the same brickwall arrangement of independently sampled 2 qubit Haar random unitaries. While the latter is not a bonafide Haar random state, it retains a strong randomising nature that exceeds the dual unitary example, in which case the random gates only act as single qubit operations. The task considered is the sinusoidal input NARMA task of order 6, for it to have sufficient memory requirements. The gates used in constructing the dual unitary reservoir have an entangling power given by $e_P(U) = 0.3$. The said comparison can be found in Fig. \ref{Figure_haar_vs_dual}.
	
	In the displayed result we compare the dual unitary reservoir in a region having intermediate mixing with the randomising circuit composed of 2 qubit Haar random unitaries. The increased performance is due to the reservoir state retaining input information in the former compared to the featureless state obtained in the latter case. Interestingly both cases of reservoir prediction outperform the single and $NV$ input regression benchmarks. From this, we can also comment that for sufficiently long evolution obtained by a large multiplexing parameter $V$ the task performance would be same between system sized Haar random evolution and the structured circuit cases that we consider. 
	
\begin{figure}
		\includegraphics[width = \linewidth]{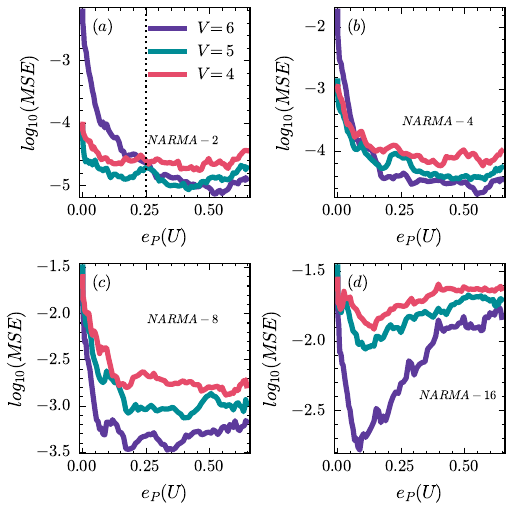}
		\caption{Variation of task performance with entangling power of gates, for increasing orders of NARMA task $(a) - (d)$, with different multiplexing appearing as parameters.}
		\label{Figure_memory_effects_multiplexing}
	\end{figure}

	Irrespective of the reservoir dynamics, increasing multiplexing generally increases the associated task performance, as the classical postprocessing step becomes more powerful with the additional nodes generated by multiplexing. However, we have also found exceptions to this general rule. For example, in Fig. \ref{Figure_memory_effects_multiplexing}, we have plotted the variation of task performance with the entangling power of gates for the dual unitary reservoir. In the different parts, we have shown the variation for increasing orders of NARMA task, each for different values of the multiplexing parameter $V$. Here, in part $(a)$ and to a smaller extent in $(b)$, it is found that in the region of low $e_P(U)$, $V = 6$ performs worse than $V = 4$ and $V = 5$. These variations are for NARMA tasks of order $2$ and $4$ respectively, which have relatively smaller memory requirements. In part $(c)$ considering NARMA task of order $8$, the variation is more typical, whereas in part $(d)$, the signature dip associated with the 'edge-of-chaos' behaviour becomes apparent, owing to the memory requirements of the said task. In the later examples, $V = 6$ consistently outperforms $V = 4$ and $V = 5$ as is the conventional expectation.
	
	The anomalous behaviour of case $(a)$ maybe explained by the greater number of nodes corresponding to $V = 6$ overfitting with respect to a poorly generated feature space, corresponding to an evolution with low mixing rate, for the given dataset. This explains the reduced performance when compared to $V = 4,5$.
	\begin{figure*}
		\includegraphics[width = \textwidth]{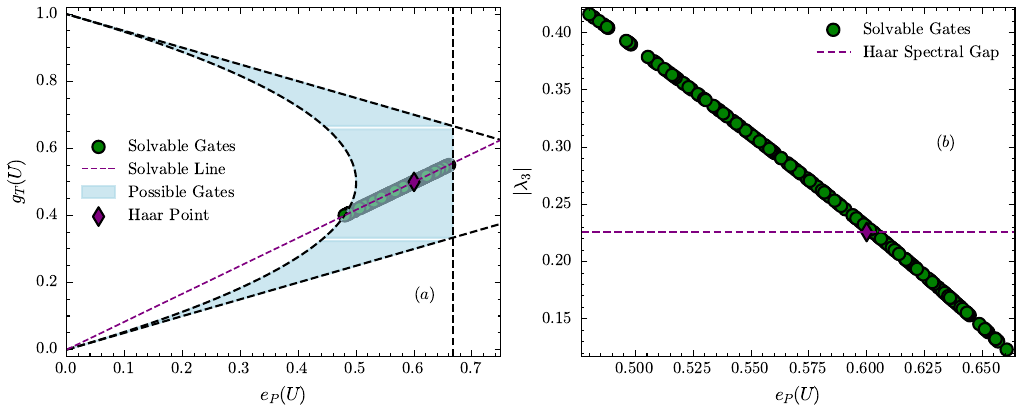}
		\caption{$(a)$ The distribution of solvable gates, each represented by a point in the $e_P-g_T$ plane, which lie upon the solvable line of $e_P(U) = 6/5 g_T(U)$. The shaded area shows the allowed region for all $2-$qubit gates. $(b)$ The variation of spectral gap $|\lambda_3|$ with entangling power for each gate shown in panel $(a)$, within the brickwall arrangement of gates for system size $N = 6$, with independently single qubit Haar random locals at each free index of gate. The spectral gap of an ensemble made up of $2-$qubit Haar random operators within the same arrangement is also shown, and has value $\sim 0.225$. When $e_P(U)$ exceed $0.6$, the average $e_P(U)$ for $2-$qubit Haar random unitaries, the spectral gap of the non Haar circuit is below that the latter, indicating greater convergence towards unitary $2-$designs.}
		\label{Figure_Solvable_Gates}
	\end{figure*}
	
	Thus, we can conclude by saying that increased multiplexing in general enhances the features generated by the reservoir unitary. Specifically for the circuit reservoir case, the light cone argument indicates a diminished rate of improvement with increasing multiplexing beyond $V \sim N$. This is somewhat observed in the task performance involving synthetic data that we consider in this work. Moreover, multiplexing over several steps would lead the reservoir state to converge close to a featureless thermal state, which would inhibit task performance by reducing the features which maybe extracted upon by evaluating the expectation value of observables. Apart from the generation of features, the task performance also depends on the efficacy of classical postprocessing, which is always benefitted by an increase in multiplexing. However, there may also be cases of overfitting as discussed above.   
	
	While having a featureless thermal state is a disadvantage in terms of information retrieval, an ensemble of unitaries showing convergence towards the Haar random distribution is an indicator of sufficiently complex dynamics. In general, a setup like this forms unitary $k$-designs, and the effect towards reservoir computing is studied in the following section. 
	
	\section{Class of Solvable Quantum Circuits}
	\label{Special}
	
	Using the Cartan decomposition, all possible $2$ qubit gates maybe represented as $U(\alpha, \beta, \gamma)$, given by:
	\begin{equation}
		U(\alpha, \beta, \gamma) = \exp \Big[ i(\alpha \sigma_x \otimes \sigma_x + \beta \sigma_y \otimes \sigma_y + \gamma \sigma_z \otimes \sigma_z)\Big],
	\end{equation} 
	under local unitary equivalence. The allowed values of parameters ($\pi/4 \ge \alpha \ge \beta \ge |\gamma| $) forms the  Weyl chamber. Thus $U$ forms the entangling non-local kernel, and is generally responsible for any dynamical properties associated with the unitary operator. 
	
	An arbitrary $2$ qubit gate having nonlocal kernel $U$ may therefore be characterised by local unitary invariant measures in the context of their dynamical properties. Important examples of said measures include entangling power $(e_P(U))$ and gate typicality $(g_T(U))$. The entangling power has been widely used in previous sections, and quantifies the average entanglement generated when the given operator acts on an ensemble of separable Haar random initial states. In contrast gate typicality quantifies the nonlocal property of a gate when compared to a typical random unitary. While statistical in nature, both quantities take a closed form upon averaging, which can be given as \cite{BhargaviJonnadula_1, BhargaviJonnadula_2}:
	\begin{equation}
		e_P(U) = \frac{E(U) + E(US) - E(S)}{E(S)}
	\end{equation}
	\begin{equation}
		g_T(U) = \frac{E(U) - E(US) + E(S)}{E(S)}
	\end{equation} 
	
	In the above, $S$ is the 2 qubit SWAP gate. $E(A)$ denotes operator entanglement of the unitary $\hat{A}$, and quantifies the nonlocal complexity of the 2 qubit unitary operator. It is obtained by Schmidt coefficients resulting from the decomposition of $\hat{A}$ in terms of orthonormal single qubit gates \cite{Zanardi_Operator_entanglement, Nielsen_resource}.
	\begin{align}
		A = \sum_{i = 0}^{3} \sqrt{\gamma_i} X_i \otimes Y_i, \& \\ 
		E(A) = 1 - \frac{1}{2^4}\sum_{i = 0}^{3} \gamma_i^2
	\end{align}  
	
	All possible $2$ qubit gates $U$, are characterised by the ordered pair $\{e_P(U), g_T(U)\}$, and occupy a section of the $e_P-g_T$ coordinate system, as shown in Fig. \ref{Figure_Solvable_Gates} $(a)$. In Ref. \cite{more_global_randomness} a subclass of the generic unitary gates is put forward, under which all the eigenvalues and eigenvectors of the second moment operator $M_{\nu}$ can be obtained for a random circuit ensemble described as $\nu$, subject to satisfying a solvability condition. The second moment operator is defined as:
	\begin{equation}
		M_{\nu} = \int U^{\otimes 2} \otimes \bar{U}^{\otimes 2} d\nu(U).
	\end{equation} 
	A given $2$ qubit operator, with the nonlocal kernel $U$ is 'solvable', if it follows the condition \cite{more_global_randomness}:
	\begin{equation}
		\frac{e_P(U)}{g_T(U)} = \frac{e_H}{g_H},
	\end{equation}
	where $e_H$ and $g_H$ are the average values of entangling power and gate typicality for the $2$ qubit Haar random measure, taking the values $0.6$ and $0.5$ respectively for the qubit case. Thus, the gates satisfying solvability condition are on the line $e_P(U) =  6/5 g_T(U)$ in the $e_P-g_T$ coordinates, as shown in Fig. \ref{Figure_Solvable_Gates} $(a)$. 
	
	Including independently sampled Haar random local unitaries at each free index of a solvable gate and constructing a brickwall circuit allows us to write the effective averaged gate obtained over all ensemble members. This averaged gate can be written in terms of the entangling power ($e_P(U)$) and gate typicality ($g_T(U)$). 	
	
	With the said arrangement of random gates, averaging over all the independently sampled Haar random local unitaries reduces the second moment operator in the permutation symmetric subspace, given by $span \{ \ket{I}, \ket{S} \}^{\otimes N}$, with the kets denoting operators mapped as states. In this subspace the dimension of $M_{\nu}$ is given as $2^N \times 2^N$, which is still tractable for a reasonable number of qubits ($N \sim 16$). Thus, including independent Haar random significantly simplifies the problem, which otherwise is highly numerically intractable, with matrix sizes scaling as $2^{2N}$. 
	
	In the permutation symmetric subspace, the averaged $2$ qubit operator is given as:
	\begin{equation}
		W = 
		\begin{bmatrix}
			1 & 0 & 0 & 0 \\
			a & b & c & a \\
			a & c & b & a \\
			0 & 0 & 0 & 1 \\
		\end{bmatrix}
	\end{equation}
	where, the terms $a,b \,\,\ \& \,\,\ c$ are given in terms of $e_P(U)$ and $g_T(U)$, as specified above \cite{more_global_randomness}:
	\begin{align}
		a = 2/3 \cdot e_P(U) \\
		b = 1 - 5/6 \cdot e_P(U) - g_T(U) \\
		c = g_T(U) - 5/6 \cdot e_P(U)
	\end{align}
	Embedding the operator $W$ in the same structure as the random circuit leads to the eigenspectrum of $M_{\nu}$ \cite{more_global_randomness}.
	
	The eigenvalues $M_{\nu}$ are obtained, and sorted in a descending order of magnitude given as $|\lambda_1| \ge |\lambda_2| \ge |\lambda_3| ... \ge |\lambda_{2^N}| $. With $M_{\nu}$ being the second moment operator the first two eigenvalues are trivial, and given by unity. The value of $|\lambda_3|$ carries important physical information, and describes the convergence rate of the given circuit to a $2$-unitary design. For example, an ensemble of random circuits, where $|\lambda_3|$ also approaches unity will show much poor convergence to the $2$-unitary design compared to an ensemble with a lower value of $|\lambda_3|$, for the same depth in terms of the applied gates. Alternatively, the latter would require a much smaller depth to form a $2$-design, compared to the former. 
	
	For solvable gates, additionally satisfying $e_P(U) > e_H (= 0.6)$, the value of $|\lambda_3|$ is necessarily below that compared to a circuit made up 2 qubit Haar random gates, in the same arrangement. This implies a faster convergence to unitary $2-$designs compared to the latter. Our results underline these observations, and are presented in Fig. \ref{Figure_Solvable_Gates} $(b)$. 
	
	For a Cartan kernel $U(\alpha, \beta, \gamma)$ satisfying the solvability condition, the following must also be satisfied \cite{more_global_randomness}:
	\begin{equation}
		f(\alpha, \beta) + f(\beta, \gamma) + f(\gamma, \alpha) = 0
		\label{solvable_eq}
	\end{equation}
	where $f(x,y)$ is given as:
	\begin{equation}
		f(x,y) = sin^2(2x)\Big( cos^2(2y) - 3/5\Big).
	\end{equation}
	Hence, to obtain a number of gates following solvability (nearly $300$ for our case), we randomly generate the parameters $\alpha$ and $\beta$ within the Weyl chamber, and then find the value of $\gamma$ using the Eq. \ref{solvable_eq}. This approach includes several instances where no solution exists for $\gamma$, although with sufficient number of repeated trials, we can indeed obtain a large collection of solvable gates.
	
	Once this collection of gates is obtained, compliance to the solvability condition is verified by plotting each operator upon the $e_P-g_T$ plane, and is found to lie on the solvable line of $e_P(U) = 6/5 g_T(U)$. With this collection of solvable gates, we shall explore the variation of $|\lambda_3|$ with $e_P(U)$ within the brickwall arrangement of gates, for system size $N = 6$. The results of this variation are shown in Fig. \ref{Figure_Solvable_Gates} $(b)$. When the entangling power of a chosen solvable gate is greater than $0.6$, the value of $|\lambda_3|$ reduces beyond the value of $\sim 0.225$, the latter being the same obtained for a brickwall circuit made up of 2 qubit random gates.
	\begin{figure}
		\includegraphics[width = \linewidth]{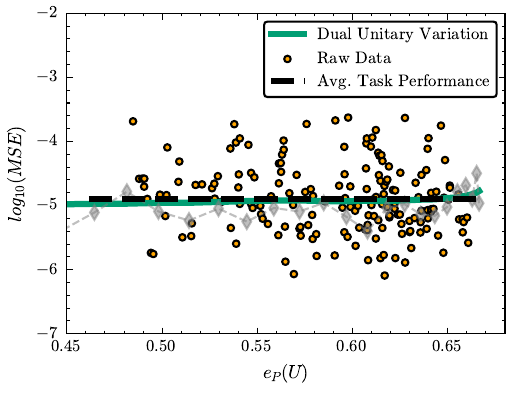}
		\caption{Variation of task performance for the Mackey-Glass dataset versus entangling power of gates $e_P(U)$ for the class of solvable gates. Although the data has high fluctuations, the task performance for solvable gates matches the dual-unitary case on average indicating similar ergodic features.}
		\label{Figure_Sovable_Gates_Performance}
	\end{figure}
	
	We shall now discuss the application of brickwall circuit with gates satisfying solvability as a reservoir and evaluate the resulting task performance. For this, we shall again plot the variation of MSE with the entangling power of gates as an ergodic indicator for the underlying circuit, given its effect on $|\lambda_3|$. Reservoir performance is benchmarked on the same Mackey-Glass dataset as in case of the previous example involving brickwall dual unitary circuits, with multiplexing parameter taken as $V = 6$. A motivation for evaluating the class of solvable circuit is to study reservoir performance under a controlled convergence to unitary designs versus the case of dual unitaries which are known to be maximally chaotic. 
	
	Another difference in the present implementation compared to the dual unitary case is the distribution of local single qubit random gates between the two implementations. In case of the dual unitary implementation, the local random gates at each index of the unitary operator are independently sampled, and this composite gate is then repeated throughout the circuit, resulting in a Floquet setup. Compared to that, the implementation for solvable gates involve different and independently sampled locals around the gate each time the operator is applied. The different configurations of local random operators both lead to tractable metrics for gauging the ergodic behaviour of the relevant circuit model.
	
	The results for task performance are shown in Fig. \ref{Figure_Sovable_Gates_Performance}. We can see that the overall performance remains largely unchanged with respect to increasing values entangling power even as it exceeds $0.6$ into the regime with faster convergence to $2$-designs. Also, the raw data has high fluctuations for different circuit implementations constructed out of gates with similar values of $e_P(U)$, when compared to the dual unitary variation. This high variation is attributed to the fact that measures related to the second moment operator is in fact an ensemble property involving a collection of circuits. However, the actual implementation in terms of a circuit for the reservoir computing application involves a single instance of the said ensemble which is then used for a time series prediction task. Hence, the strong fluctuations are present between different implementations. 
	
	It is also important to note that the average task performance across the different entangling powers for the solvable circuit matches the same evaluated for the dual unitary case with equal multiplexing, as shown in Fig. \ref{Figure_MSE_vs_mixing_MG}. This implies that for different dynamical examples which directly converge to a high degree thermalisation, the associated task performance for the reservoir setup is quite identical. This was also alluded to in the previous section, when comparing the dual unitary example with a brickwall circuit composed of independently sample $2$ qubit gates. In the previously described case dual unitary reservoirs edged ahead, and we stated that for a long circuit depth, the performance would converge taking the example of a NARMA-6 task with sinusoidal inputs. The comparision at present centres on the single-step ahead Mackey-Glass forecasting, which requires retaining the memory of said dynamics. Thus, upon reaching a reservoir state that is sufficiently randomising, the autonomous task performance, on average, tends to saturate at a fixed value. 
	
	For the solvable case, the similarity in task performance across the different quantum circuit setups is reasoned to be present via the convergence to increasing orders of unitary $k$-designs, beyond the case of $k = 2$, when taking into account the associated multiplexing given by $V$ to be same as $N$.
	\begin{figure*}
		\includegraphics[width = \linewidth]{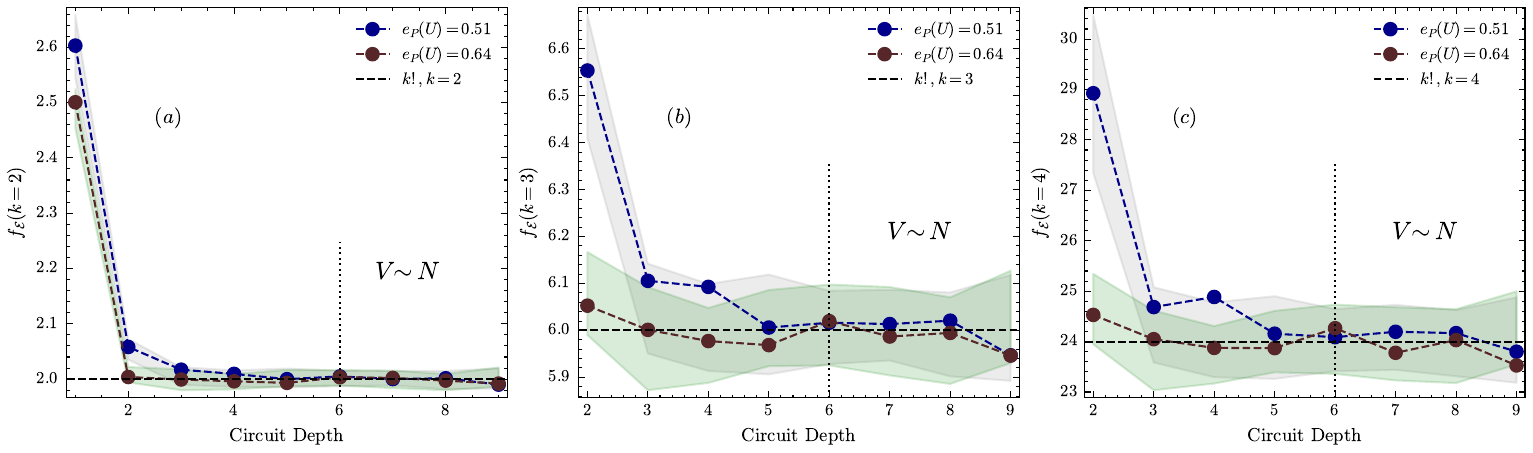}
		\caption{Convergence of frame potential for brickwall circuit made of solvable gates with entangling powers lower and higher than $e_H = 0.6$, along with circuit depth for $k = 2,3 \,\, \& \,\, 4$. For each implementation, convergence of the case with higher entangling power exceeds that of Haar, which in turn exceeds the example with lower entangling power. The shaded regions show the standard deviation observed under a total of 50 repeated implementations involving separate local unitaries.}
		\label{Figure_Solvable_Frame_Potential}
	\end{figure*}  
	
	In order to validate the convergence of the solvable circuit to increasing orders of unitary $k$-designs, we plot the frame potential for an ensemble of circuits within the solvable class for examples of higher and lower entangling power. The frame potential for an ensemble $\mathcal{E}$ in the $k$th moment is defined as follows:
	\begin{equation}
		f_{\mathcal{E}}(k) = \frac{1}{|\mathcal{E}|^2} \sum_{U,V \in \mathcal{E}} \big[ \tr(U^\dag V) \big]^{2k}
	\end{equation}
	
	For an ensemble of circuits $\mathcal{E}$ forming a $k$-design, the value of frame potential matches that of $k!$ \cite{frame_potential1, design_codes_scott}.
	\begin{equation}
		f_{\mathcal{E}}(k) = k!
	\end{equation}
	Using these metrics, we find that for solvable gates under the condition $e_P(U) > 0.6$ the convergence to unitary $k$-designs is indeed faster for all orders of increasing $k$. Although in context of task performance, we see that even in case of slower convergence the circuit approaches $k$-design in $V \sim N$ multiplexing steps. This primarily explains the mechanism behind similar average performance across entangling powers.
	
	The results of this analysis are shown in Fig. \ref{Figure_Solvable_Frame_Potential}. We find that the circuit setup shows a rapid convergence to the class of $k-$designs, for intermediate values of $k = 2,3$ and $4$. To obtain the results, we have generated a $1000$ realisations of the random circuit having a depth of $10$ time steps, for each value of the considered entangling power. This entire implementation is also repeated $50$ times to evaluate the deviations between ensemble implementations. For all repetition, the higher $e_P(U)$ show faster convergence, followed the example with gates having lower $e_P(U)$. The same is observed for all the considered orders. Moreover, the deviation in convergence observed for different realisations is small as shown by the shaded regions in Fig. \ref{Figure_Solvable_Frame_Potential} which indicate the standard deviation in the values of frame potential over the $50$ different realisations.  
	
	We now relate the task performance of the circuit class of solvable gates with the frame potential indicators. As shown in Fig. \ref{Figure_Solvable_Frame_Potential}, for $N=6$ and multiplexing chosen as $V\sim N$, circuits composed of solvable gates converge to $k$-designs for increasing orders of $k$, irrespective of the entangling power being tuned for faster convergence. The primary distinction between different values of $e_P(U)$ is therefore the rate of convergence, which is faster for gates with larger entangling power, as expected from the variation of $|\lambda_3|$ with entangling power discussed earlier in this section and shown in Fig. \ref{Figure_Solvable_Gates}. This validates the notion that for solvable gates forming $k$-designs at depth $V$ irrespective of entangling power, the task performance remains the same on average across different values of $e_P(U)$, with the aforementioned fluctuations. 
	
	In either examples that we consider, the high scrambling nature proves to be useful as it allows for enhanced spreading of information through the system and hence rapidly constructs a feature space for the input data injected into the system, leading to high expressivity of the reservoir states, at the cost of a lower memory retention. When compared to the case of Hamiltonian dynamics, where it is possible to have reservoirs with ergodicity breaking mechanisms, the memory retention for circuit reservoirs is lower. This shortcoming still allows for an efficient implementation of the quantum reservoir procedure, requiring a small depth of circuit operations per input data point, and hence maybe useful in context of NISQ era devices. In terms of resources, the class of brickwall circuits that we consider are economical compared to the standard case of system sized Haar random unitaries, which used as a standard example for simulating thermalising quantum dynamics \cite{Haferkamp2022, Knill1995Approximation, Oszmaniec2021EpsilonNets}. This is due to the requirement of single or two qubit random unitaries for the circuit examples we consider. 
	
	Thus, an implementation of the kind we have considered should be expected to give good task performance with relatively lower computational resources compared to Hamiltonian reservoirs, owing to a structural thermalising nature which yields peak performance at multiplexing $V \sim N$, versus $V \sim 20$ for the latter case \cite{Cindrak_Krylov_Observability}. This efficiency gain assumes that the task at hand has a limited timescale and hence smaller memory requirements.

	\section{Summary and Conclusions}
	\label{Summary}

	In this work, we have investigated the use of structured quantum circuits as reservoirs for the purpose of temporal information processing. The reservoir is constructed from a brickwall arrangement of two qubit gates, where the input time series is sequentially injected into the first qubit and the resulting dynamics is sampled throughout the system by local single qubit measurements. The measured expectation values, collected over a multiplexing depth $V$, form the effective feature space used by the linear readout layer. This setup allows us to study how the dynamical properties of the system sized circuit, controlled by the nature of its constituent two qubit gates, influences reservoir performance for a specific task.
	
	We primarily consider the class of dual unitary circuits as the quantum reservoir, where the ergodic nature of the reservoir unitary is directly governed by the entangling power of the constituent two qubit gates. In this setup, modifying the entangling power directly changes the mixing nature of the dynamics. To account for the role of single qubit local unitary operators we fix them to have the maximal mixing possible, a measure which is monotonically dependent on entangling power. In this dynamical situation the saturation of Krylov complexity follows the mixing nature of the circuit, and increases for larger mixing. With this observation, we are led to conclude that with increased mixing, the input information is spread into a larger feature space, thereby being indicative of superior predictive performance. Since the nature of Krylov vectors converges for large trotter steps towards maximally chaotic dynamics \cite{Claeys_Circuit_Krylov, Wisniacki_Krylov_Claeys}, different examples of the same brickwall arrangement of gates would have similar Krylov state behaviour indicating their usefulness for reservoir computing. With the geometrical structure of the circuit fixed in the brickwall arrangement we also obtain results for the constituent gates being $2$ qubit Haar random along with the solvable set of gates.
	
	The task performance of the circuit reservoir is validated using synthetic datasets, where we use the NARMA task to gauge the memory retention capacity of the reservoir along with the chaotic Mackey-Glass series to evaluate the forecasting capability of the reservoir. Here, the impact of reservoir dynamics towards task performance is estimated by comparing the accuracy of prediction against appropriate classical regression benchmarks. The primary benchmark in this regard is the single input linear regression, as the same is used for making the final prediction by optimising over expectation value of observables extracted from the reservoir. Another, more stringent benchmark is against a delayed input linear regression, which takes the previous $NV$ inputs of the time series for predicting the output, having $NV + 1$ nodes to optimise against. In all cases, we find that the reservoir significantly outperforms the former, while the latter generally requires greater multiplexing, especially when the input is randomised. The second benchmark constitutes an explicit memory baseline, which measures the performance of linear regression when it is provided with a recent history of inputs to determine the output. For this reason, specifically when the input has no temporal relationship, the linear regression can outperform the reservoir in certain cases. For the reservoir to be better than the delayed input linear regression, it requires a larger feature space, which maybe generated by increased multiplexing. Note that while the second linear regression benchmark is explicitly provided with the recent history of the time series variation, the reservoir on the other hand constructs the features by the process of multiplexing, being given only a single input. 
	
	We also specifically study the role of multiplexing towards reservoir performance. Increased multiplexing generally leads to improved task performance, combining the effects of a larger feature space generated by reservoir dynamics along with a larger set of parameters to optimise against for the final classical postprocessing step. However, the role of reservoir dynamics is somewhat diminished after multiplexing $O(N)$ times, by which stage the input information is spread throughout the system by means of the brickwall circuits. As a result, increasing the multiplexing beyond $V \sim N$ leads to diminishing improvements in prediction accuracy. Another factor is that multiplexing several times leads the reservoir state to resemble the feature less thermal state, from which it is difficult to extract information that would aid task prediction.
	
	The reservoir performance following near complete randomisation is studied by using $2$-qubit Haar random gates in the brickwall arrangement, along with the class of solvable gates. The latter shows a controlled convergence towards unitary $2$-designs and higher upon varying the entangling power of the constituent gates. We find that using the NARMA task, the performance of circuit involving Haar random gates is always lower than that of the dual unitary case. For the Mackey-Glass forecasting, which involves constructing features of the time series by sampled observations, the task performance of solvable gates matches the dual unitary case on average. This indicates that upon sufficient thermalisation the average task performance obtained from the reservoir, across different implementations, tends to converge. Moreover, this point of convergence represents lower prediction accuracy than what can be obtained by the dual unitary dynamics, at optimal operation. This observation is consistent with the traditional 'edge-of-chaos' understanding in reservoir computing for optimal task performance. Further, we also observe a enhancement of task performance when using the dual unitary reservoir at small, nonzero values of mixing, which further supports the 'edge-of-chaos' regime as optimal for the case of circuit reservoirs. Thus, while increasing the ergodicity associated with the reservoir unitary leads to a larger feature space, it also reduces the memory capacity of the reservoir, which is why the best performance is found by counterbalancing between the two opposing requirements. 
	
	In summary, our work evaluates quantum circuits as reservoirs, and in the process sheds light on the mechanisms associated with unitary evolution that are relevant for reservoir computing applications. We find the structured examples of quantum circuits like dual unitary or solvable gates, owing to their scrambling nature, yield their peak performance in relatively smaller circuit depth corresponding to the multiplexing requirement of $V \sim N$. This represents an implementation efficiency compared to Hamiltonian reservoirs, where peak performance is obtained around $V \sim 20$ \cite{Cindrak_Krylov_Observability}. 
	
	Future directions to explore include analysing circuits closer to the 'edge-of-chaos' by incorporating effects of ergodicity breaking while also exploring the real quantum device implementations with noise models and measurement induced backaction \cite{Duals_scars, Duals_measurement, ETH_Quantum_Ckts, Sourav_MIPT, Ergodicity_breaking_cell_auto}. Another direction to explore are the effects of non-stabiliserness on reservoir performance, particularly comparing the task performance for the class of Clifford dual unitaries in the strong and weak scrambling limits \cite{Claeys_Clifford_Duals}. These directions shall allow a deeper analysis towards reservoir probing for characterising quantum dynamics \cite{res_probe, res_probe_pt}. Finally, the analysis could incorporate real datasets for testing the feasibility of quantum reservoir computing for temporal data processing \cite{CM_Res1}. 
	
	Moreover, since the completion of this work, another study has investigated the effectiveness of dual unitary circuits as a platform for quantum reservoir computing \cite{Claeys_dual_reservoir}. The authors describe the role of operator dynamics, particularly soliton propagation and operator spreading, in generating memory and nonlinear processing capabilities. Furthermore, they demonstrate that dual unitary circuits can mitigate exponential concentration of measurement outcomes, thereby improving robustness against finite shot noise. Our findings in this work are consistent with the results presented in Ref. \cite{Claeys_dual_reservoir} towards the entangling nature of the circuit.
	
	\section*{Acknowledgment}
	We wish to acknowledge the resources of supercomputing facility 'Param Shivay' at IIT(BHU), which were used to generate the results presented in this work. SA further acknowledges funding support from the National Quantum Mission, an initiative of the Department of Science and Technology, Govt. of India and the support provided by the Foundation for QC Innovation (FQCI), DST-NQM T-Hub at IISc Bengaluru, in facilitating this work.
	
	\section*{Data Availability}
	The data that support the findings of this article are openly available \cite{github_data_reference_reservoir}.
	
	\bibliography{circuit_reservoir}
	\appendix
	\section{Mixing Rate and role of local unitaries in dual-unitary circuits}
\label{Mixing_Rate}

For completeness, we briefly summarize the elements of the dual-unitary mixing formalism required in the present work. The full derivation and notation were developed in our previous work, Ref. \cite{literature1}, to which we refer the reader for details.

For a dual-unitary two-qubit gate \(U\), correlations along the light cone are governed by the single-site quantum channel
\begin{equation}
	\mathcal{M}_{\nu}(a)=\frac{1}{q}\operatorname{tr}_1\!\left[U^\dagger(a\otimes I)U\right].
\end{equation}

Upon vectorization, this channel is represented by a superoperator \(M_{\nu}(U)\). Its largest eigenvalue is unity, while the magnitude of the largest nontrivial eigenvalue, \(|\lambda_1|\), determines the mixing rate,
\begin{equation}
	\mu_1=-\ln |\lambda_1|.
\end{equation}

The spectrum of \(M_+(U')\), for \(U'=(u_1\otimes u_2)U(v_1\otimes v_2)\), depends on the local-unitary dressing even though \(e_P(U)\) remains invariant. In the present work, we therefore select, for each value of \(e_P(U)\), the locally equivalent gate that maximizes \(\mu_1\). The corresponding maximal-mixing relation used in the main text follows from Ref. \cite{literature1}. A more comprehensive discussion regarding dual unitary dynamics, correlation function and the mixing rate can be found in Refs. \cite{Aravinda_Sir_PRR, Dual_Unitary1}.

\section{Krylov Complexity using Circuit Operator}
\label{Krylov_basics}

In this section, we shall specify the methods used to generate the Krylov basis, given an initial operator mapped to its corresponding state vector and the reservoir unitary. The successive time evolved versions of the state vector are obtained by repeated application of the single step circuit unitary $\mathcal{U}$, following which we make use of Arnoldi iterations which involve the explicit orthogonalisation of each new vector with respect to all previously created vectors. Although not necessary for the case involving dual-unitary circuits, we still include the step of orthogonalising each new time evolved vector with all previously generated vectors for numerical stability.

In order to find the Krylov basis in the space of time evolved operators, generated by the   for numerical stability. The steps followed, for an initial operator $\hat{O}$ are given as:
\begin{itemize}
	\item $\ket{\mathcal{O}_0} = \ket{\hat{O}}/||\hat{O}||$. We also have $b_0 = ||\hat{O}|| = \langle \hat{O} | \hat{O} \rangle^{1/2}$.
	\item for $t \ge 1$, $\ket{\hat{A}_t} = \ket{\mathcal{U^{\dag}}^t \hat{O}_{0} \mathcal{U}^t}$. For numerical stability, we still orthogonalise $\ket{\hat{A}_t}$ with respect to all previous basis vectors: \\ $\ket{\hat{A}_t} = \ket{\hat{A}_t} - \sum_{m = 0}^{t-1} \ket{\hat{O}_m}\langle{\hat{O}_m}|\hat{A}_t\rangle$.
	\item $b_t = \langle \hat{A}_t | \hat{A}_t \rangle^{1/2}$. These are the Arnoldi coefficients, which must be stored after each iteration. They take the same value as parameter $\beta_n$ for the maximally scrambling case.
	\item $\ket{\mathcal{O}_t} = \frac{1}{b_t}\ket{\hat{A}_t}$, if $b_t \neq 0$, else: Terminate. These are the Krylov basis vectors. These must also be stored at the end of each iteration.
\end{itemize}
Therefore, at the end of all iterations we end up with the set of Krylov basis $\{ \mathcal{O}_t \}$ and the Arnoldi coefficients $\{ b_t \}$. Note that $\ket{\hat{X}}$ represents the operator $\hat{X}$ as a member of the Hilbert space that is formed of all operators and the corresponding inner product. 

We take $t$ till $K = d^2 - d + 1$, which is the dimension of maximal Krylov space when evaluated via the traditional Liouville operator $\mathcal{L}(\hat{O}) = [\hat{H},\hat{O}]$. For the case of dual-unitaries the iterations may comfortably exceed this bound and lead to nearly linearly independent state, with the orthogonal nature deviating slightly as reported in the Section \ref{Brickwall_Reservoirs}. In other words, the condition $b_t = 0$ is never reached fully. In comparison for examples of Hamiltonian evolution that Krylov space is saturated within a few iterations.

\section{Evaluating Frame Potential for Clifford Circuits}

\begin{figure}
	\includegraphics[width = \linewidth]{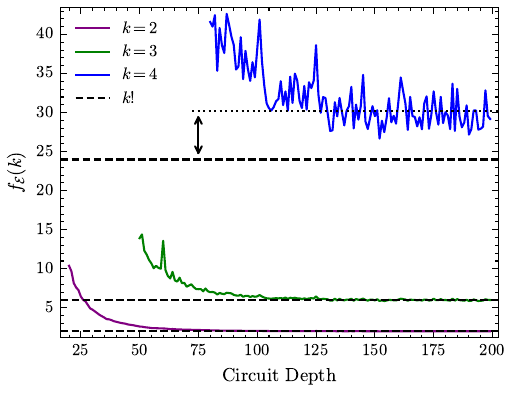}
	\caption{Variation of the frame potential with depth for an example of Clifford circuit. The frame potential converges to $k!$ for $k = 2,3$, whereas the convergence remains adrift of the value $k = 4$. This validates the known result that Clifford circuits for 2 and 3 designs, but cannot form a $k = 4$ design.}
	\label{Figure_Clifford_Frame_Potential}
\end{figure}

We shall validate the frame-potential measure for a class of Clifford circuits. The circuit  constructed by randomly choosing neighrbouring sites in a circuit of $N = 6$ qubits, followed by randomly choosing one of the following two-qubit gates \cite{Haferkamp2}:
\begin{equation}
	G = \{ H \otimes \mathbb{I}, S \otimes \mathbb{I}, S^3 \otimes \mathbb{I}, CNOT \}
\end{equation} 
Repeating the process $n$ number of times gives us a circuit of depth $n$. Since the circuit is made up of Clifford gates, it would lead to the formation of a maximum of unitary $3-$design. Further, it is known that for $k = 4$ there would be no convergence in this setup \cite{Zhu_Clifford_4design}. The variation of frame potential with the circuit depth is shown in Fig. \ref{Figure_Clifford_Frame_Potential}.

The said convergence is confirmed by plotting the frame-potential $f_{\mathcal{E}}(k)$ with circuit depth, for an ensemble of 1000 circuits made up of Clifford gates, shown in Fig. \ref{Figure_Clifford_Frame_Potential}. For this simple implementation, the resulting ensemble forms a $k = 2,3$ design in around 70 to 100 random application of said gates, observed by the convergence of $f_{\mathcal{E}}(k)$ to $k!$. For $k = 4$ the convergence saturates at value around 30, which is higher that the corresponding value of $k! = 24$. 

Here, following the manner in which the circuit construction is defined, a single layer of depth includes just a solitary Clifford gate. This explains the long circuit depth required even for the convergence to a unitary $1$-design. Moreover, with this result the convergence shown in Fig. \ref{Figure_Solvable_Frame_Potential} confirms the emergence of higher order designs involving non-Clifford resources, which are examples of solvable gates along with the independently sampled single-qubit Haar random gates at each space-time point.

\end{document}